\documentclass{pasa}

\usepackage[authoryear]{natbib}
\bibpunct{(}{)}{;}{a}{}{,}
\begin{document}

\title[Optimal SPH Initial Conditions]{Generating Optimal Initial
  Conditions for Smoothed Particle Hydrodynamics Simulations}

\author[Diehl et al.]{S. Diehl$^{1,2}$, G. Rockefeller$^{2}$,
  C.~L. Fryer$^{2}$, D. Riethmiller$^{3}$ \& T. S. Statler$^{3,4}$\\
  \affil{$^1$Nuclear \& Particle Physics, Astrophysics \& Cosmology
    Group (T-2), Los Alamos National Laboratory, P.O. Box 1663, Los
    Alamos, NM 87545}
  \affil{$^2$Computational Physics \& Methods (CCS-2), Los Alamos
    National Laboratory, P.O. Box 1663, Los Alamos, NM 87545}
  \affil{$^3$Astrophysical Institute, Ohio University, Athens, OH
    45701}
  \affil{$^4$National Science Foundation, USA}}

\begin{abstract}
  We review existing SPH setup methods and outline their advantages,
  limitations and drawbacks.  We present a new method for constructing
  initial conditions for smoothed particle hydrodynamics (SPH)
  simulations, which may also be of interest for N-body simulations,
  and demonstrate this method on a number of applications.  This new
  method is inspired by adaptive binning techniques using weighted
  Voronoi tesselations.  Particles are placed and iteratively moved
  based on their proximity to neighboring particles and the desired
  spatial resolution.  This new method can satisfy arbitrarily complex
  spatial resolution requirements.
\end{abstract}


\maketitle


\section{Introduction}

Smoothed particle hydrodynamics (SPH) is a Lagrangian hydrodynamics
modeling technique developed independently by \citet{LucySPH} and
\citet{GingoldSPH}. In this grid-less technique, fluid elements are
represented by individual particles that act according to hydrodynamic
flow equations. SPH has been used to model a wide variety of
astrophysical phenomena, including star formation
\citep[e.g.,][]{MonaghanStarFormation, BonnellStarFormation,
  WhitworthSPHJeans}, planet formation
\citep[e.g.,][]{BenzPlanetFormation, MayerPlanetFormation,
  NelsonPlanetFormation}, cosmology \citep[e.g.,][]{NavarroCosmology2,
  NavarroCosmology, SpringelGadget2}, stellar collisions
\citep[e.g.,][]{BenzStellarCollisions, RasioStellarCollisions,
  DaviesStellarCollisions, DaviesStellarCollisions2}, stellar mergers
\citep[e.g.,][]{RasioStellarMerger, TermanStellarMerger,
  DaviesStellarMerger, RosswogNSmerger, LeeStellarMerger,
  FryerStellarMerger, YoonDDSPH, DiehlCommonEnvelope, DiehlDDmerger},
gas dynamics in the Galactic center \citep[e.g.,][]{GabeGalCenter2,
  GabeGalcenter, CuadraGalCenter}, galaxy mergers
\citep[e.g.,][]{1995ApJ...448...41H, 1996ApJ...464..641M,
  ThakarDiskGals, CoxSSmerger, KhalatyanFeedback} and supernovae
\citep{FryerSN3d, HungerfordGammaray, FryerMixing}.

Each SPH particle $i$ has an associated size $h_i$, its so-called
smoothing length. Fluid properties such as temperature or density are
smoothed according to a smoothing function $W$, which is referred to
as the SPH kernel. The most commonly used kernel functions are cubic
splines \citep{MonaghanSPH} that are non-zero only within two
smoothing lengths of the particle.\footnote{The GADGET astrophysical
  SPH code defines the kernel to be non-zero from $r=0$ to $1h$
  instead of $0$ to $2h$ \citep{springel01}.} Fluid properties at the
location of a particle can then be calculated as a linear combination
of the contributions from all neighboring particles. Thus, it is
essential for this technique to start out with initial conditions
whose interpolation properties are as accurate as possible. In
addition, the initial particle setup should be as close as possible to
a configuration that would arise by itself in an SPH simulation.

Due to their well-known interpolation properties and ease of
construction, the simplest setup schemes often arrange particles on a
lattice.  While there are many lattice configurations that could in
principle be used to produce SPH initial conditions, we focus on three
popular configurations---a simple cubic lattice, a cubic close-packed
lattice, and a hexagonal close-packed lattice. The simplest such
arrangement (and one of the most popular) is the cubic lattice
configuration, which has been shown to be an unstable equilibrium
configuration and has strong preferred directions along the x, y and z
axes \citep{morris96,LombardiMixing}.  Cubic close-packed and
hexagonal close-packed lattices represent the two optimal and most
efficient ways to pack spheres of equal sizes; they are stable against
random perturbation and thus much preferred to a simple cubic lattice
\citep{MonaghanSPH}. We also include a comparison with a new
configuration method based on a quaquaversal tiling of space that has
recently been suggested for quasi-random initial conditions of
cosmological N-body simulations \citep{HansenQuaqua}.

To avoid geometrical effects, initial conditions are often perturbed
and then relaxed into a stable configuration by applying a dampening
force that is proportional to but directed against the particle
velocities \citep[e.g.,][]{RosswogWDdisruption}. While this additional
step before calculation is perfectly acceptable to produce low-noise
initial conditions, it is computationally expensive and usually only
viable for static initial conditions, as the net forces on the whole
set of particles should vanish. In addition, there is no way to
guarantee the exact configuration into which the particles will settle
at the end.

Another major problem in setting up initial conditions for SPH is that
many astrophysical simulations require very large dynamic ranges in
density. In simulations of, e.g., the accretion flow in binary mass
transfer, the convective region in core-collapse supernova engine
models, or interactions between supernova remnants or stellar winds
and the interstellar medium, the resolution requirements may not trace
the mass, and a range of particle masses may be required to model the
system.  Large ranges of particle masses in SPH are undesirable and
care must be taken when using a range of particle masses.  However,
setting up arbitrary initial conditions with a spatially varying
resolution is an unsolved problem so far; the few previously proposed
solutions have only been applicable in spherical symmetry
\citep[e.g.,][]{FryerMixing, RosswogWDdisruption}.

In this paper, we propose a solution to this problem inspired by
weighted Voronoi tesselations (WVTs) and present a new method to set
up SPH initial conditions with arbitrary, spatially varying resolution
requirements. We describe requirements for an optimal setup technique
in \S\ref{s.requirements} and then review and compare existing popular
particle setup methods in \S\ref{s.methods}. To the best of our
knowledge, this is the first comprehensive comparison of SPH setup
techniques, despite the known importance of initial conditions for SPH
simulations. We introduce our new setup in \S\ref{s.wvt} and
demonstrate its capabilities with examples in \S\ref{s.examples}. In
\S\ref{s.comparison} we quantitatively compare this new setup method
to existing techniques.


\section{Requirements for an Optimal Particle Configuration Method}
\label{s.requirements}

A method for generating initial SPH particle configurations should
fulfill the following key requirements:

\paragraph*{Isotropy.}

The resulting particle configuration should be locally and globally
isotropic, i.e., it should not impose any particular preferred
direction at any location in the simulation domain. The main reason
for this requirement is the fact that shocks moving along a perfectly
aligned string of SPH particles behave differently than in other
directions \citep{HerantDirtyTricks}. In addition, spatially
correlated density perturbations can excite modes along these
preferred directions.

\paragraph*{High Interpolation Accuracy.}

The setup should be locally uniform to minimize noise in the density
interpolation. Ideally, for a uniform resolution, the interpolation
accuracy should be comparable to that of perfectly uniform lattice
configurations. This interpolation accuracy should also worsen for
non-uniform particle configurations. Any deviations should also be
isotropic and have no preferred directions, in order not to excite
non-physical modes in the simulation domain. This requirement is
equivalent to enforcing a low particle noise.

\paragraph*{Versatility.}

The ideal method should be able to reproduce any spatial configuration
and should not impose any requirement for symmetry. In particular,
this requires the method to work with interpolation of a tabulated
data set and not require analytical solutions.

\paragraph*{Ease of Use.}

Ideally, the algorithm should either be publicly available as a
stand-alone routine or be easy to implement on top of any existing SPH
code.


\section{Popular Particle Setup Methods}
\label{s.methods}

\begin{figure*}
  \begin{center}
    \includegraphics[width=0.32\textwidth]{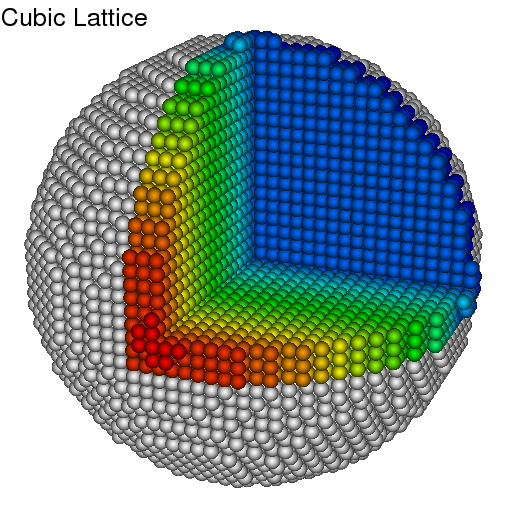}
    \includegraphics[width=0.32\textwidth]{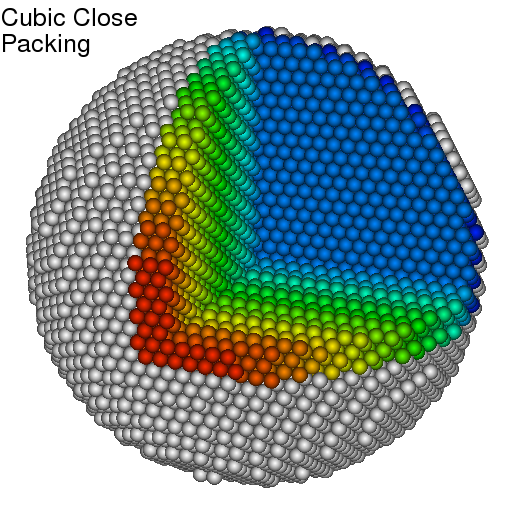}
    \includegraphics[width=0.32\textwidth]{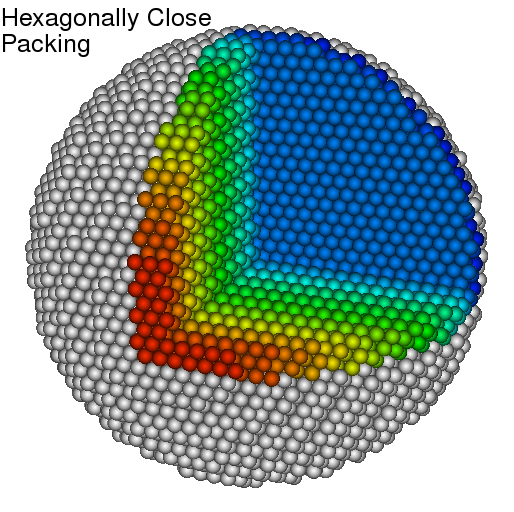}
    \includegraphics[width=0.32\textwidth]{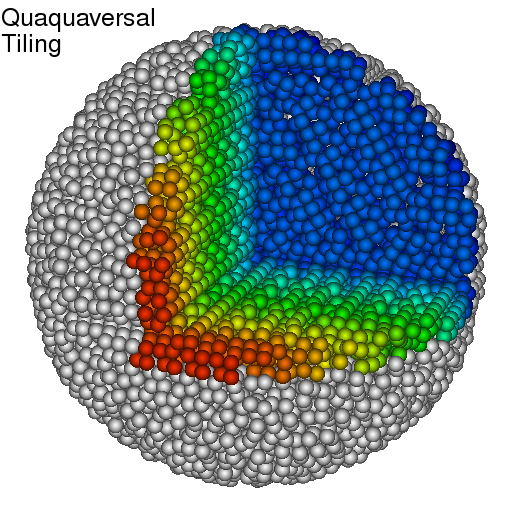}
    \includegraphics[width=0.32\textwidth]{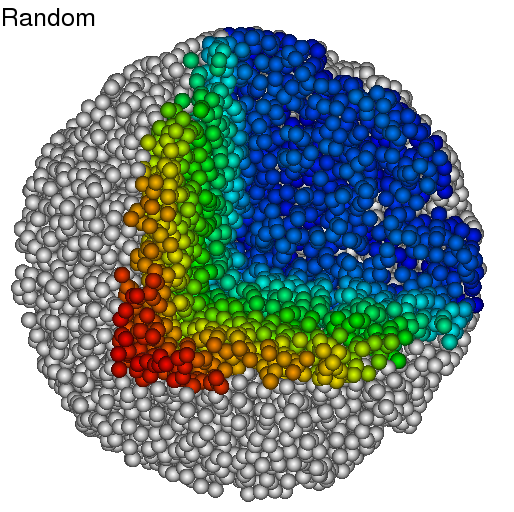}
    \includegraphics[width=0.32\textwidth]{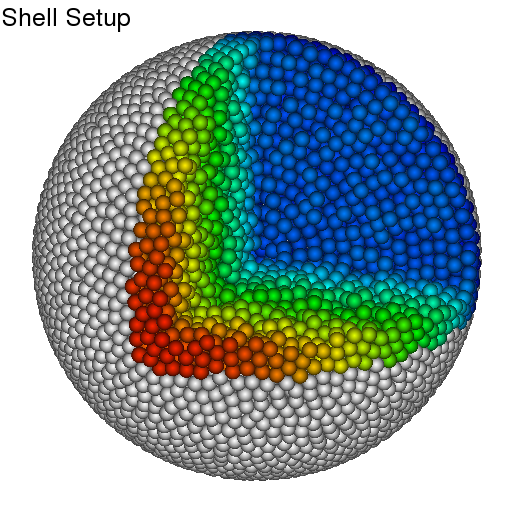}
    \includegraphics[width=0.32\textwidth]{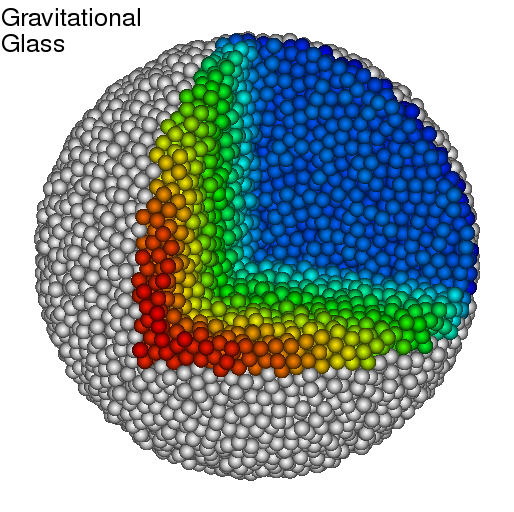}
    \includegraphics[width=0.32\textwidth]{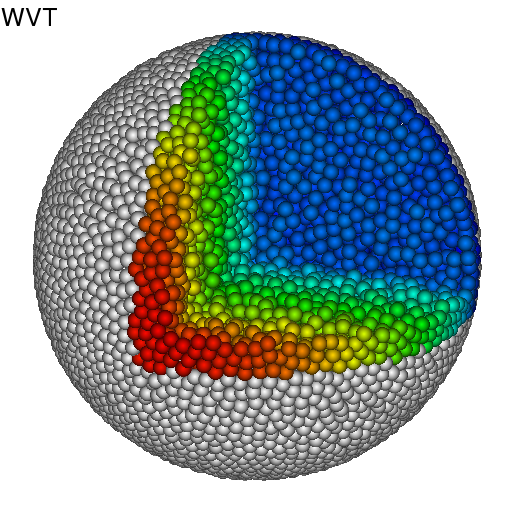}
  \end{center}
  \caption{Popular configurations for setting up spatially uniform SPH
    initial conditions.  From the top left corner to the bottom right:
    cubic lattice, cubic close packing, hexagonal close packing,
    quaquaversal tiling, random configuration, concentrical shells,
    gravitational glass, and the new WVT approach. All examples
    contain approximately the same number of particles in the sphere
    ($22,000$). One quadrant of the sphere is cut out to allow a view
    into the inner configuration. Colors change along the z-axis
    simply to show depth.}
  \label{f.popularconf_uniform}
\end{figure*}

\begin{figure*}
  \begin{center}
    \includegraphics[width=0.32\textwidth]{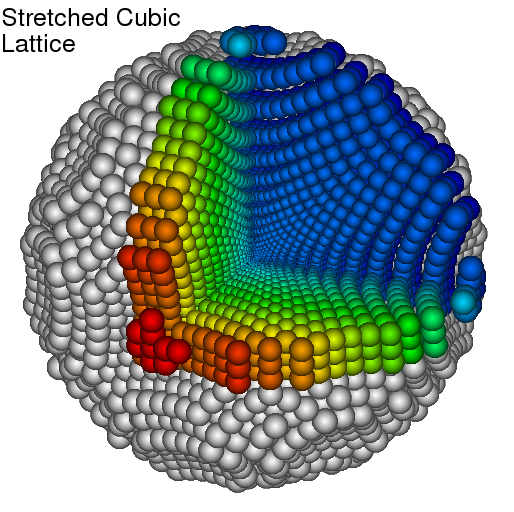}
    \includegraphics[width=0.32\textwidth]{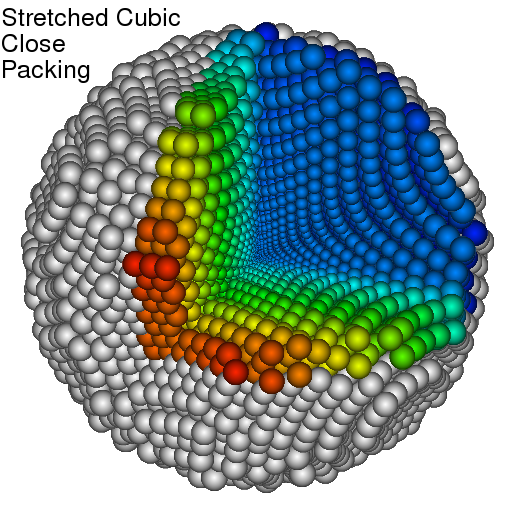}
    \includegraphics[width=0.32\textwidth]{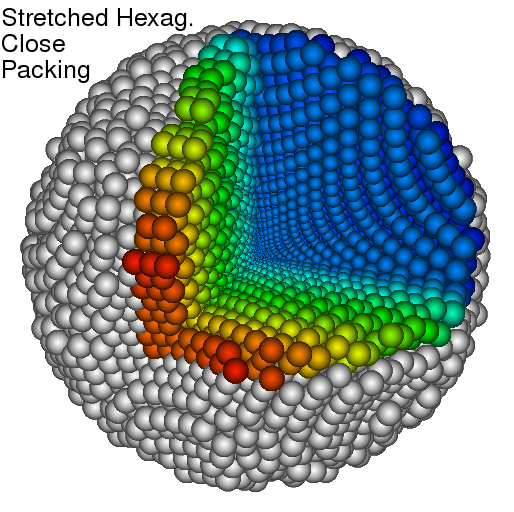}
    \includegraphics[width=0.32\textwidth]{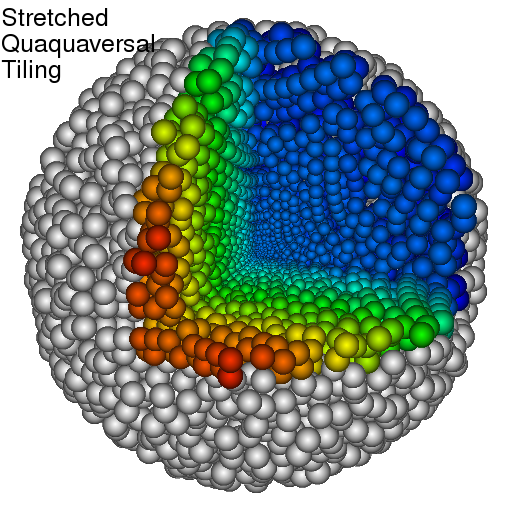}
    \includegraphics[width=0.32\textwidth]{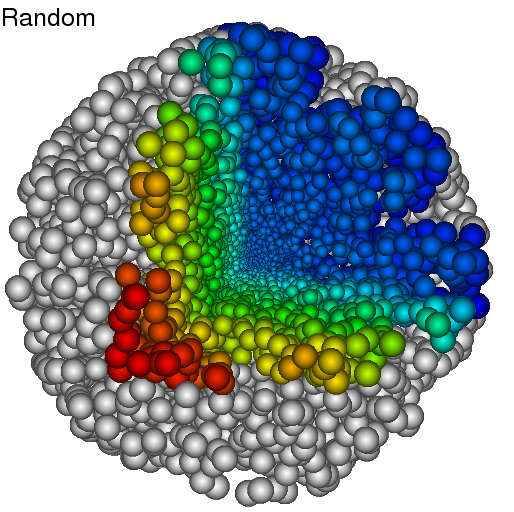}
    \includegraphics[width=0.32\textwidth]{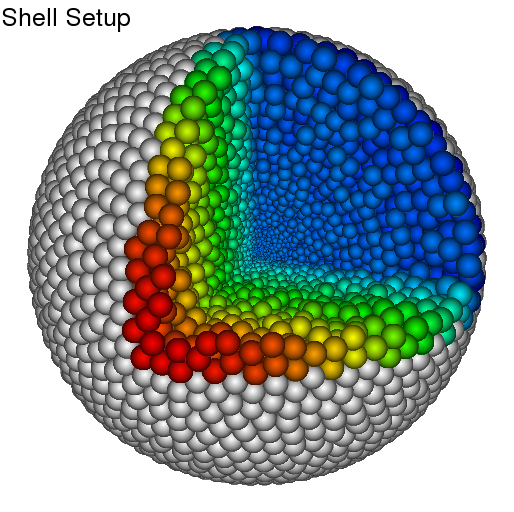}
    \includegraphics[width=0.32\textwidth]{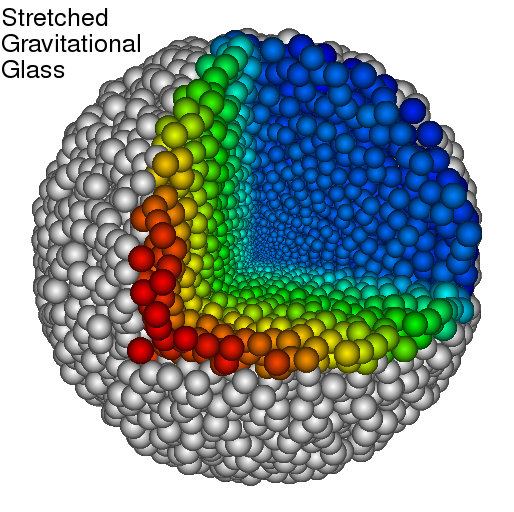}
    \includegraphics[width=0.32\textwidth]{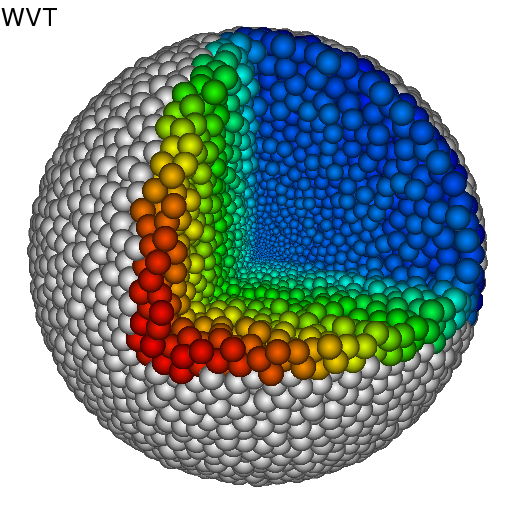}
  \end{center}
  \caption{Popular configurations for setting up spatially adaptive
    SPH initial conditions.  From the top left corner to the bottom
    right: stretched cubic lattice, stretched cubic close packing,
    stretched hexagonal close packing, stretched quaquaversal tiling,
    random configuration, concentrical shell setup, stretched
    gravitational glass, and the new WVT approach. All examples
    contain approximately the same number of particles in the sphere
    ($22,000$), and the particles' sizes reflect the desired particle
    spacing. One quadrant of the sphere is cut out to allow a view
    into the inner configuration. Colors change along the z-axis
    simply to show depth.}
  \label{f.popularconf_adaptive}
\end{figure*}

Since the invention of the SPH technique, many different methods have
been employed to set up initial conditions in multiple dimensions. In
this section, we summarize all particle setup methods known to us, or
that have been described in the
literature. Figure~\ref{f.popularconf_uniform} shows a simple
comparison of results of arranging approximately 22,000 particles with
equal smoothing lengths in a sphere with each method.
Figure~\ref{f.popularconf_adaptive} shows a similar comparison for
spatially-adaptive configurations, where the smoothing length varies
across the domain (from smallest at the center, to largest at the
outer boundary). \S\ref{s.spatially_uniform} describes the methods
that are limited to producing configurations in which all particles
have the same smoothing length; \S\ref{s.spatially_adaptive} describes
the methods that can be used when smoothing lengths are not uniform.
Readers who are already familar with these methods or who are just
interested in details of our method may examine the figures or skip
this section entirely.

\subsection{Spatially Uniform Distributions}\label{s.spatially_uniform}

The following methods are capable of generating spatially uniform
particle configurations.

\paragraph*{Cubic Lattice.}

Probably the simplest and fastest way to set up a uniform particle
distribution is to arrange them on a cubic lattice. This method
received early widespread use in both SPH \citep{MonaghanSPH} and
N-body simulations \citep{EfstathiouNbody}. One of the obvious
problems with this method is that it has very pronounced preferred
directions along the x, y, and z axes and their diagonals, as can
easily be seen in the upper left example in
Figure~\ref{f.popularconf_uniform}. In addition, the cubic lattice
structure is not a stable equilibrium configuration when the particles
are perturbed \citep{morris96,LombardiMixing}, as there are other more
compact particle configurations that are energetically favorable, such
as cubic or hexagonal close-packed arrangements.

\paragraph*{Cubic Close-Packed Lattice.}

A more compact lattice structure is produced when one inserts an
additional particle into the center of each of the six faces of the
cubes in the cubic lattice. This results in the well-studied cubic
close packing (CCP) configuration, also known as face-centered cubic
close packing (Figure~\ref{f.popularconf_uniform}, top center
panel). This configuration is one of the optimal ways to pack uniform
spheres together, with a packing density of $74\%$. Similar to the
cubic lattice, it has the problem of having preferred directions along
the principal axes and diagonals of the lattice, and along multiple
other planes in which particles are arranged in a hexagonal
grid. However these preferred direction are much less pronounced than
for the cubic lattice configuration. The simplest way to construct
this configuration is to start with a plane with spheres in a
hexagonal configuration (plane A), and lay on top another such plane
(B) so that the spheres fit into the gaps created by the layer of
spheres from the lower plane, filling half the gaps. The third such
plane (C) will then be oriented in a way to fill the other half of the
gaps of plane A, while at the same time fitting into the gaps of plane
B. This pattern is then continuously repeated to produce an ABCABC
order.

\paragraph*{Hexagonal Close-Packed Lattice.}

A very similar lattice configuration is the second optimal packing
scheme of uniform spheres, hexagonal close packing (HCP), as seen in
the top right panel of Figure~\ref{f.popularconf_uniform}. HCP is
equally dense and optimal as CCP, with very similar properties. The
only difference in its construction is that instead of the ABC pattern
as in the CCP lattice, every second layer of hexagonal lattice planes
is identical, resulting in an ABAB pattern. Due to their relatively
simple implementation, close packing schemes have been utilized in
many different applications of SPH
\citep[e.g.,][]{DaviesStellarCollisions, DaviesStellarCollisions2}.

\paragraph*{Quaquaversal Tiling.}

Another lattice-like particle configuration has been introduced by
\citet{HansenQuaqua} based on a quaquaversal tiling of space
\citep{ConwayQuaqua}. Quaquaversal tiling hierarchically tiles the 3d
space into triangular prisms that are rotated about orthogonal axes
$2/3\,\pi$ and $1/2\,\pi$. Originally introduced as a way to set up
cosmological initial conditions, recent work by \citet{WhiteNbodyIC}
argues against this choice. As can be seen in the middle left panel of
Figure~\ref{f.popularconf_uniform}, this setup has many
characteristics of a grid.

\paragraph*{Gravitational Glass.}

The best way to set up cosmological initial conditions however is the
generation of a gravitational glass. This method simply reverses the
sign of gravity and lets the particles settle into an equilibrium
configuration while dampening their motion. In this paper, we use the
implementation of a gravitational glass as provided in the
publicly-available Gadget2 code
\citep{springel01,SpringelGadget2,WhiteNbodyIC}. This method is
particularly effective when used with periodic boundary conditions. A
cube with a fixed number of particles can then be replicated numerous
times to achieve a larger total number of particles in the final
configuration. While this is disadvantageous for cosmological N-body
simulations, which require a certain degree of homogeneity on all
scales, SPH only requires locally optimal configurations and is not
affected by this. Thus, a single instance of a gravitational glass can
be concatenated to effectively produce arbitrarily large particle
configurations.

\subsection{Spatially Adaptive Distributions}\label{s.spatially_adaptive}

We now discuss the subset of setup methods that are capable of
producing spatially adaptive particle distributions. To the best of
our knowledge, all of these methods have so far been employed to
create spherically symmetric configurations. Examples of each of these
methods are shown in Figure~\ref{f.popularconf_adaptive}.

\paragraph*{Random Configuration.}

The simplest option to produce a spatially adaptive particle
configuration is to distribute particles randomly according to an
underlying probability distribution. For example,
\citet{TermanStellarMerger} have used this technique in combination
with the relaxation method to create adaptive initial conditions. This
relaxation is absolutely necessary, as this method results in very
clumpy distributions with very low interpolation accuracy. We only
mention this method as it represents the starting point for our new
setup procedure described in \S\ref{s.wvt}.

\paragraph*{Stretched Lattice.}

To achieve a spatially adaptive resolution, \citet{HerantDirtyTricks}
and later \citet{RosswogWDdisruption} proposed to stretch a uniform
lattice configuration in the radial direction. With this method, each
point coordinate ${\bf r}$ of the uniform lattice is multiplied by a
radially varying scaling factor $q(r)$ to achieve the desired
spherically symmetric distribution, such that ${\bf r'}=q(r)\, {\bf
  r}$. This also implies through simple geometry that the given
distance $\delta$ between two particle on the shell with radius $r$ is
now also scaled by $q(r)$, effectively setting the resolution to
$\delta'(r')=q(r)\, \delta$.

Thus, the problem has now been reduced to figuring out how to choose
$q(r)$ to produce the desired resolution in the new stretched
coordinates, i.e. $\delta'(r')$ has to obey the differential equation
$r' \delta(r) - r \delta'(r') = 0$. While it is entirely possible to
solve this problem analytically for simple $\delta'(r')$ functions by
substituting $r'$ and solving, it is more convenient to use a more
generally applicable technique to find the root of the function
$f(r')=r' \delta(r) - r \delta'(r')$, with its derivative $df'/dr' =
\delta(r) - r\,d\delta'(r')/dr'$.  We chose the simple Newton-Raphson
technique by iterating over
$r'_{n+1}=r'_n+f(r'_n)\,[df/dr'(r'_n)]^{-1}$.  Note that $r$ is a
constant parameter in this context, as it is given by the known
position of the particle in the uniform lattice.

As this distorted lattice essentially incorporates and even aggravates
all of the undesirable characteristics of a lattice configuration, it
is essential {\it not} to use this stretched lattice configuration
directly, but rather to relax the resulting configuration before using
it in an actual SPH simulation \citep{RosswogWDdisruption}.

\paragraph*{Stretched Glass.}

Instead of radially stretching or compressing a uniform lattice
configuration, it is also in principle possible to generate a
gravitational glass of uniform resolution and stretch this glass
accordingly, to avoid the strong preferred lattice symmetry axes. To
the best of our knowledge, this method has never been employed in
published simulations.

\paragraph*{Concentric Shells.}

In many supernova calculations using the Supernova SPH code (SNSPH)
\citep[e.g.,][]{FryerSN3d, HungerfordGammaray, SNSPH, FryerMixing,
  FryerConvection}, the initial conditions are set using shell
templates.  For a given particle count, such a shell template is
created by first randomly placing the particles in a shell of unit
radius.  The particles are given a repulsive force and then the entire
system is evolved until the variation in particle separation falls
below some tolerance, essentially creating a two-dimensional
gravitational glass wrapped around a sphere.  The templates are then
used to match a given spherical density profile and either a
resolution or particle mass requirement.  The spherical object is
constructed from the inside outward, each concentric shell determining
the new position of the next shell (one smoothing length above the
previous shell).  The shells are randomly rotated and placed on top of
each other so that, even if the same template is used, the setup is
random. A variant of this method had originally been proposed by
\citet{HerantDirtyTricks} but to our knowledge has never been
extensively described in the literature before.

The advantage of this technique is that the particles are placed
randomly and hence have no preferred axis.  Depending on the tolerance
set for the template creation, the resolution used, and the density
gradient, this technique can match a spherical density profile to
arbitrary precision.  For low resolution core-collapse calculations,
\citet{FryerConvection} achieved density perturbations below 3-5\%
(convection in the stellar models they were mapping argued for higher
perturbations).  For a high resolution mapping of an exploding star,
\citet{FryerMixing} limited this perturbation to below 1\%.  This
technique is tuned to spherical objects, and does not work, without
major revision, on aspherical objects.


\section{A New Approach Inspired By Weighted Voronoi
  Tesselations}
\label{s.wvt}

In this section, we describe a novel technique for generating
spatially adaptive initial conditions for SPH simulations that does
not impose any restrictions on the geometry of the desired
configuration.  This method was inspired by a two-dimensional adaptive
binning technique using weighted Voronoi tesselations developed by
\citet{DiehlWVT}, which in turn was based on previous work by
\citet{Cappellari}.

\subsection{Weighted Voronoi Tesselations}

Given a metric and a set of $k$ points ${\bf z}_{i}$, $i=1,...,k$
(referred to as ``generators'') in a given domain, a Voronoi
tesselation of the domain is a tesselation in which the $i$th region
contains all of the points closer to ${\bf z}_i$, according to the
chosen metric, than to any other generator.  A weighted Voronoi
tesselation applies a weight to the distance from each generator; a
multiplicatively weighted Voronoi tesselation simply multiplies the
distance from a given generator by its associated weight \citep[see,
  e.g.,][]{Moller}.  The boundary surface ${\bf b}$ between adjacent
regions in a multiplicatively weighted Voronoi tesselation is defined
such that the scaled distance from each generator to the surface is
equal, i.e.,
\begin{equation}
  \label{e.wvt}
  |{\bf b}-{\bf z}_i|/\delta_i=|{\bf b}-{\bf z}_j|/\delta_j,
\end{equation}
if the metric is simply Euclidean distance, and where $\delta_i$ is a
scale factor (i.e., the inverse of the weight) assigned to the $i$th
generator.

A centroidal Voronoi tesselation (CVT) is a Voronoi tesselation where
each generator coincides with the centroid of its region.  Again, each
generator can have an associated weight or scale factor, so that the
sizes of the regions in the CVT vary across the domain.

One of the most well-known algorithms for constructing a CVT is the
Lloyd algorithm \citep{Lloyd}, which alternates between constructing a
Voronoi tesselation from a set of generators, and moving each
generator to the centroid of its associated Voronoi region.  The Lloyd
algorithm is a special case of a general gradient descent approach to
minimizing the CVT energy function \citep{DuAccelCVT, LiuAccelCVT}.
It delivers monotonic, linear convergence to a CVT without the need
for step size control \citep{DuCVTReview} but requires the ability to
construct Voronoi tesselations, which is not a capability typically
associated with implementations of SPH.

\subsection{Technique}

Our adaptive setup technique arranges particles by repeatedly applying
net displacements based on proximity to neighbors and the desired
final spatial resolution, which can vary across the problem domain.
Identifying neighboring particles and iteratively accumulating and
applying pairwise displacements falls well within the normal
capabilities of SPH codes.

Although we refer to it as ``WVT'' because of the method that inspired
it \citep{DiehlWVT}, our technique does not actually construct a WVT.
The technique described below is similar to the distributed algorithm
for constructing area-centered Voronoi configurations
\citep{CortesMartinezBullo, MartinezCortesBullo}; each particle within
a limited radius contributes to the calculation of a displacement in
each iteration.

\begin{figure*}
  \begin{center}
    \includegraphics[width=0.24\textwidth]{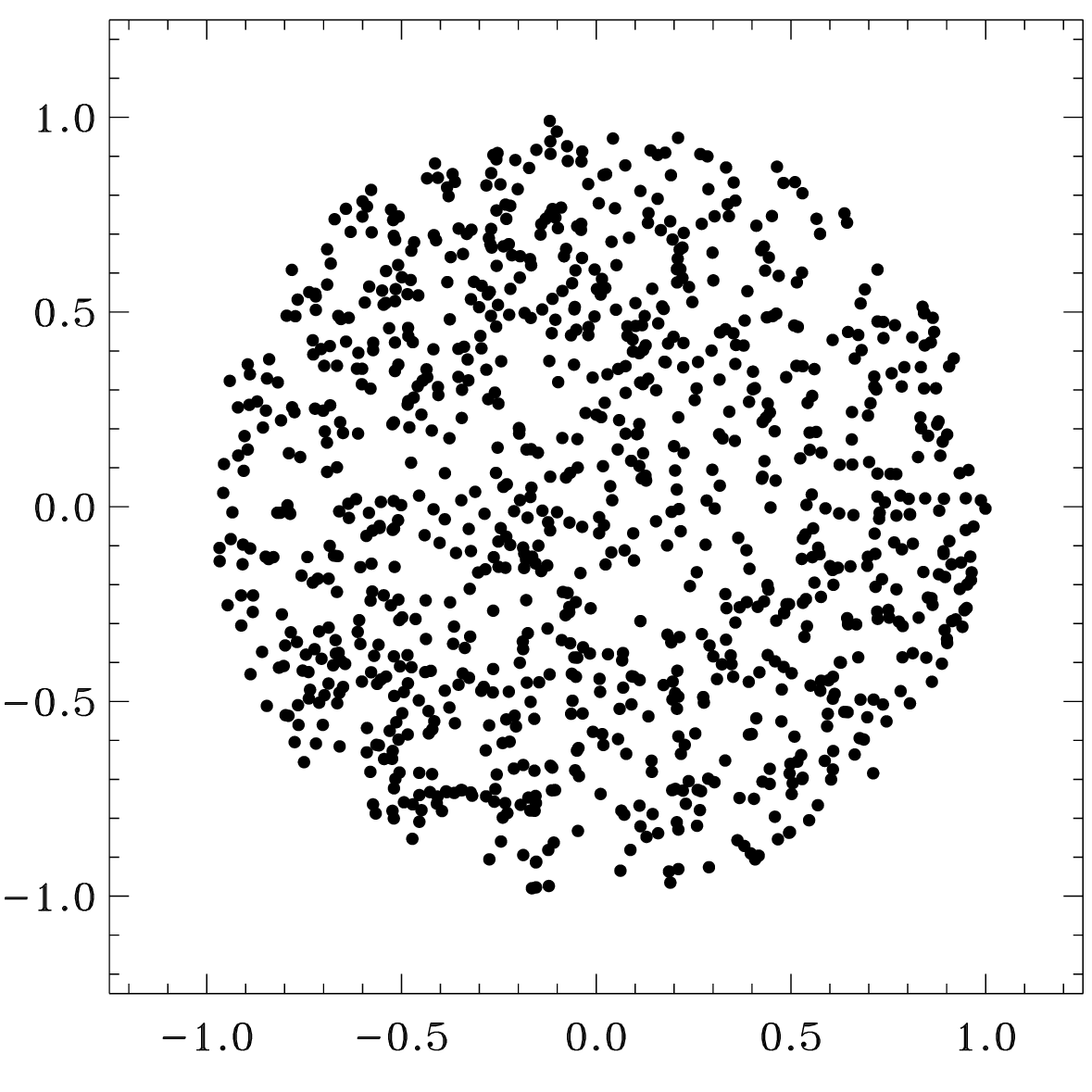}
    \includegraphics[width=0.24\textwidth]{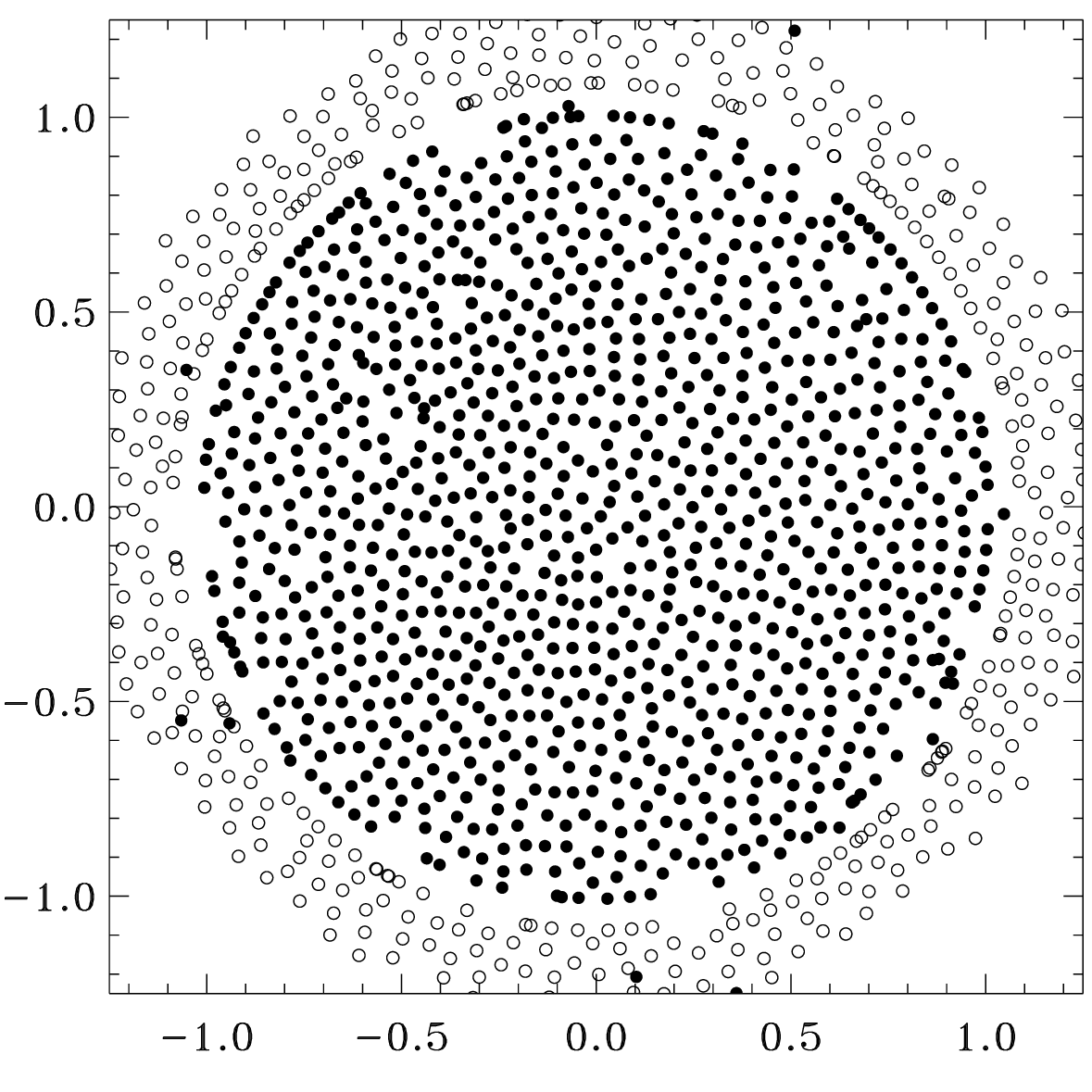}
    \includegraphics[width=0.24\textwidth]{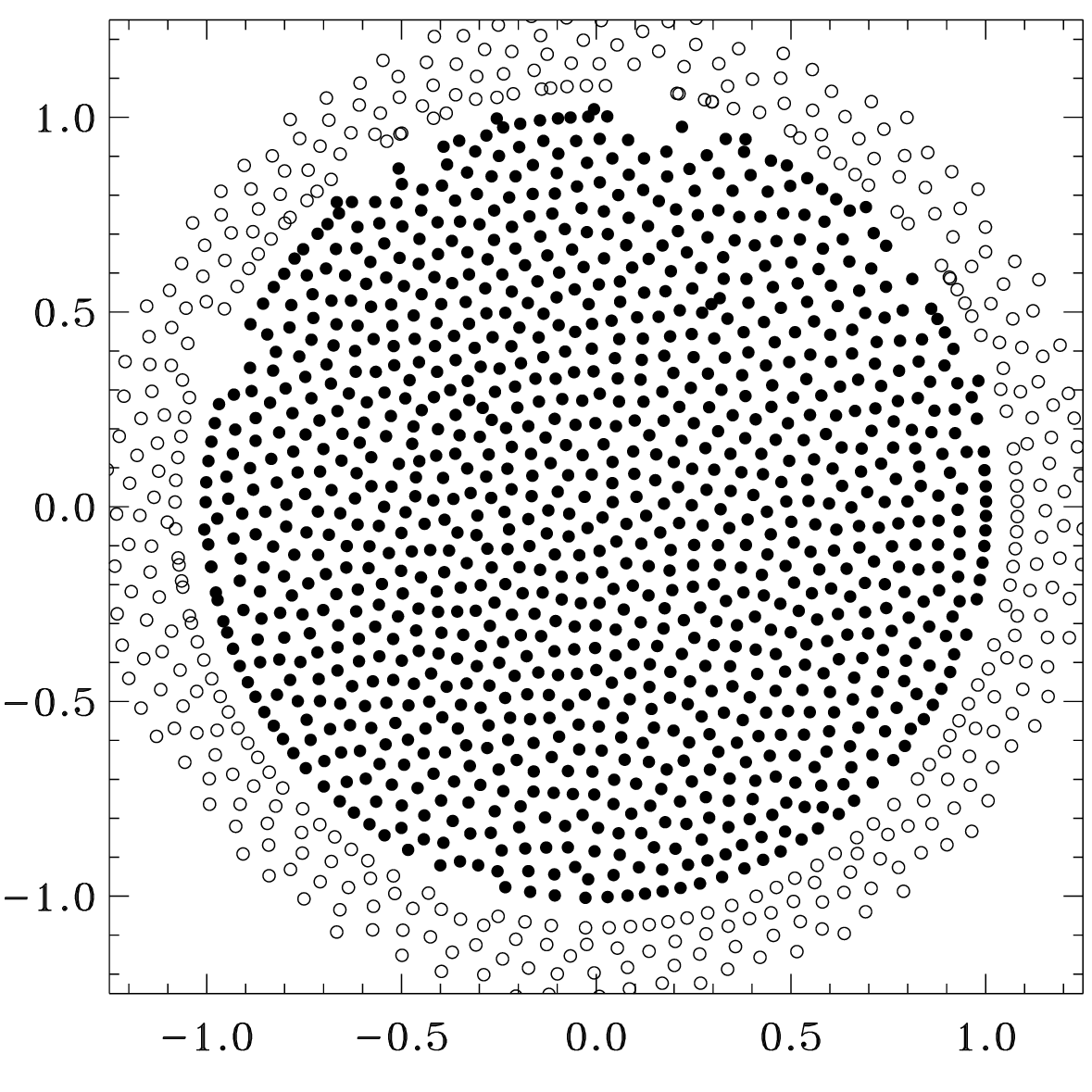}
    \includegraphics[width=0.24\textwidth]{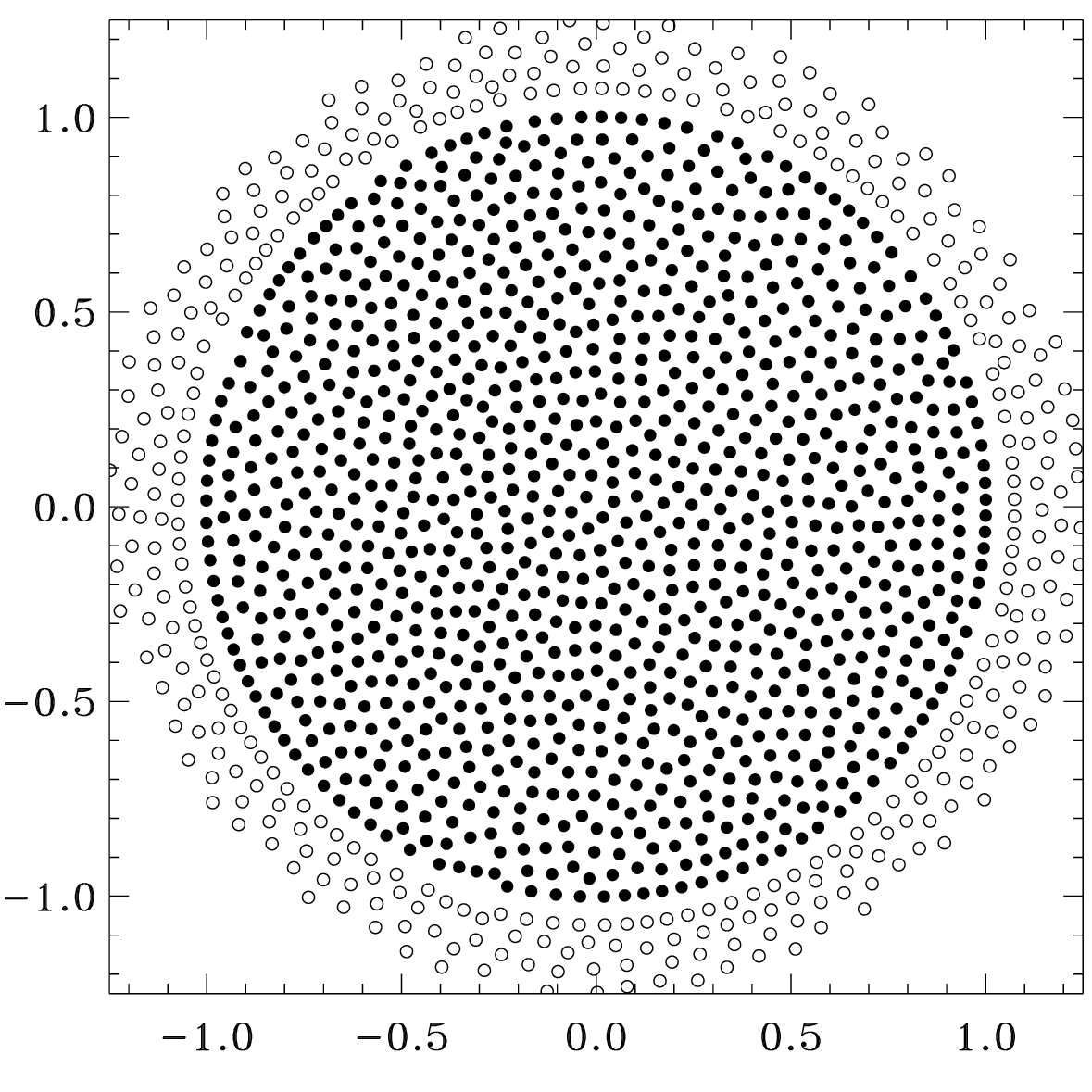}\\
    \includegraphics[width=0.24\textwidth]{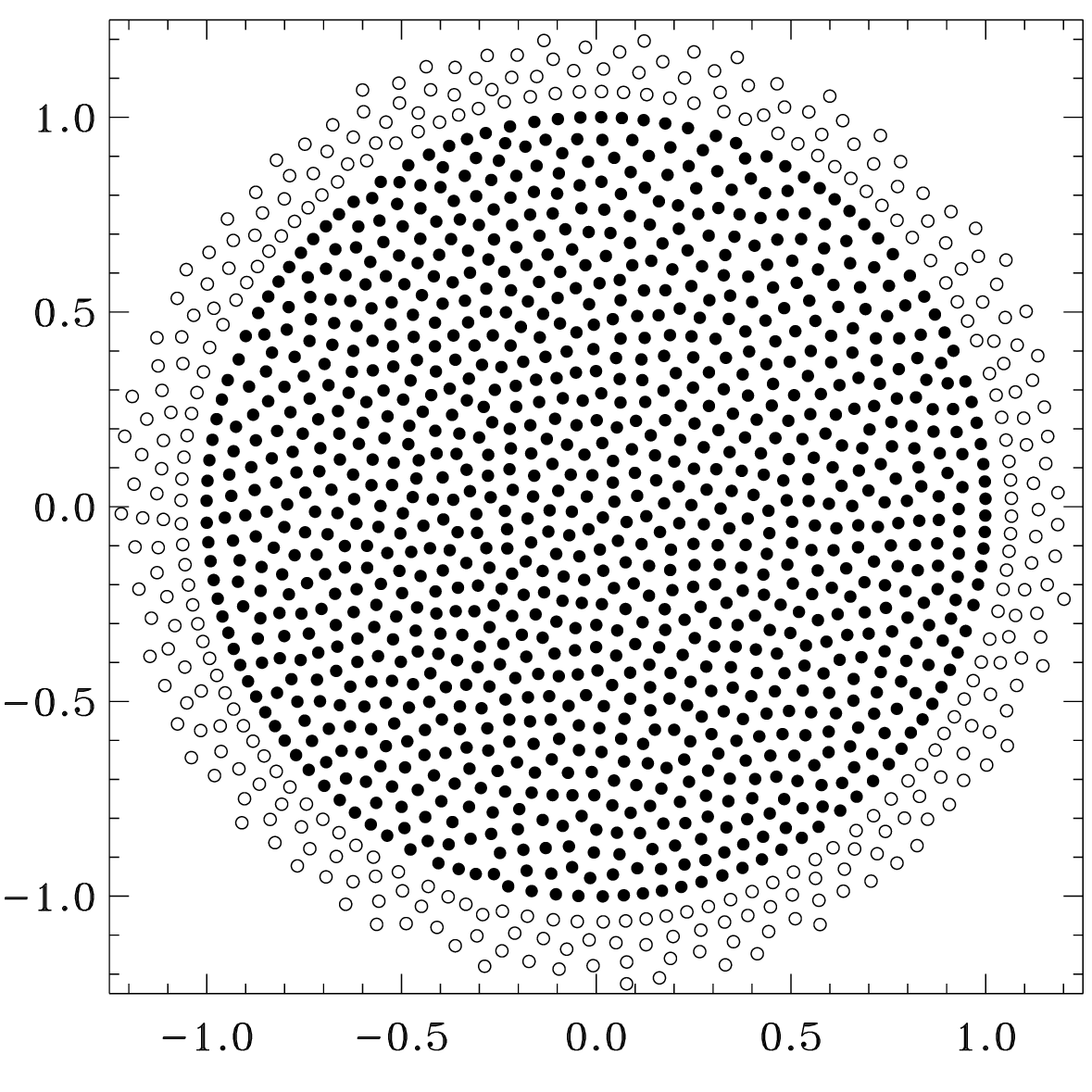}
    \includegraphics[width=0.24\textwidth]{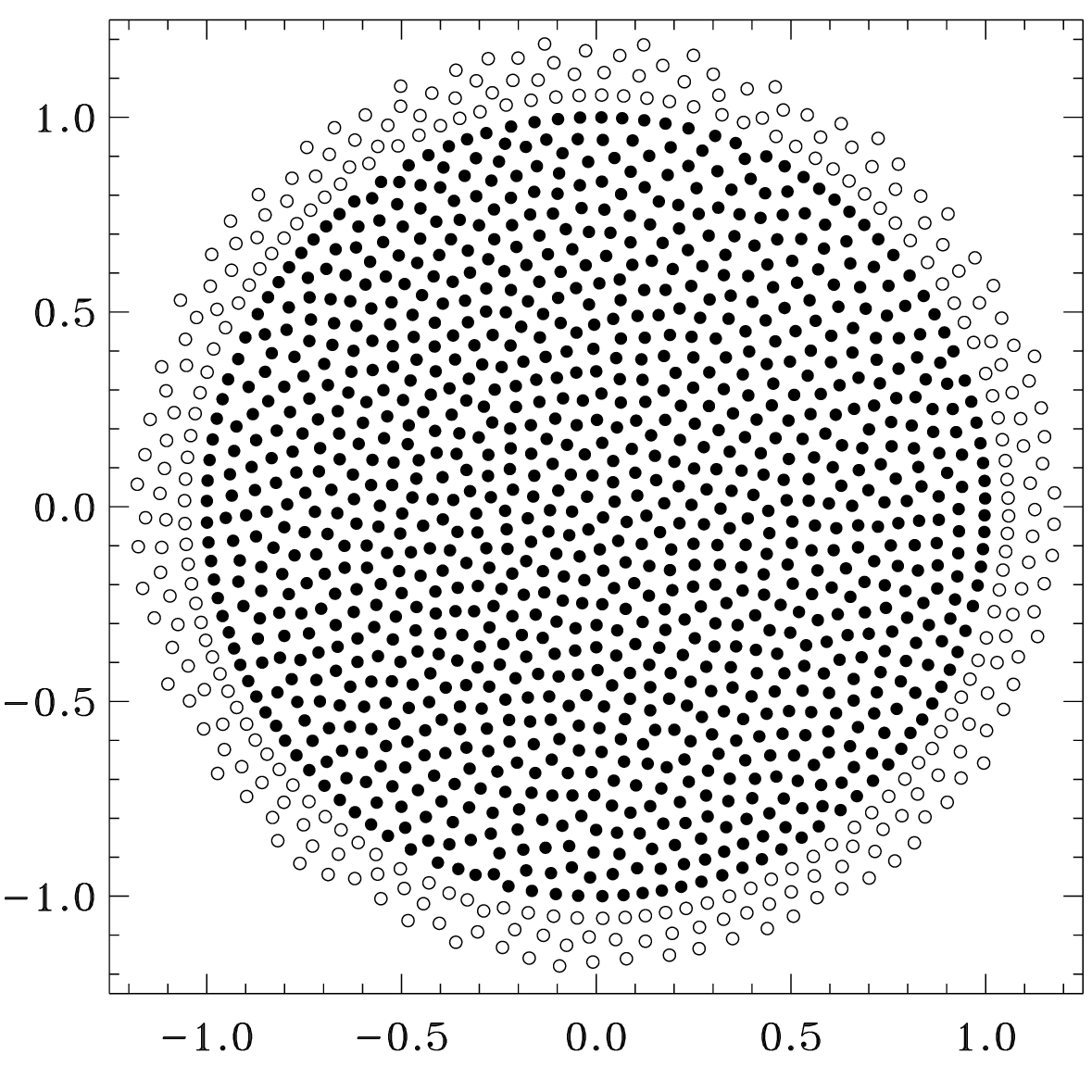}
    \includegraphics[width=0.24\textwidth]{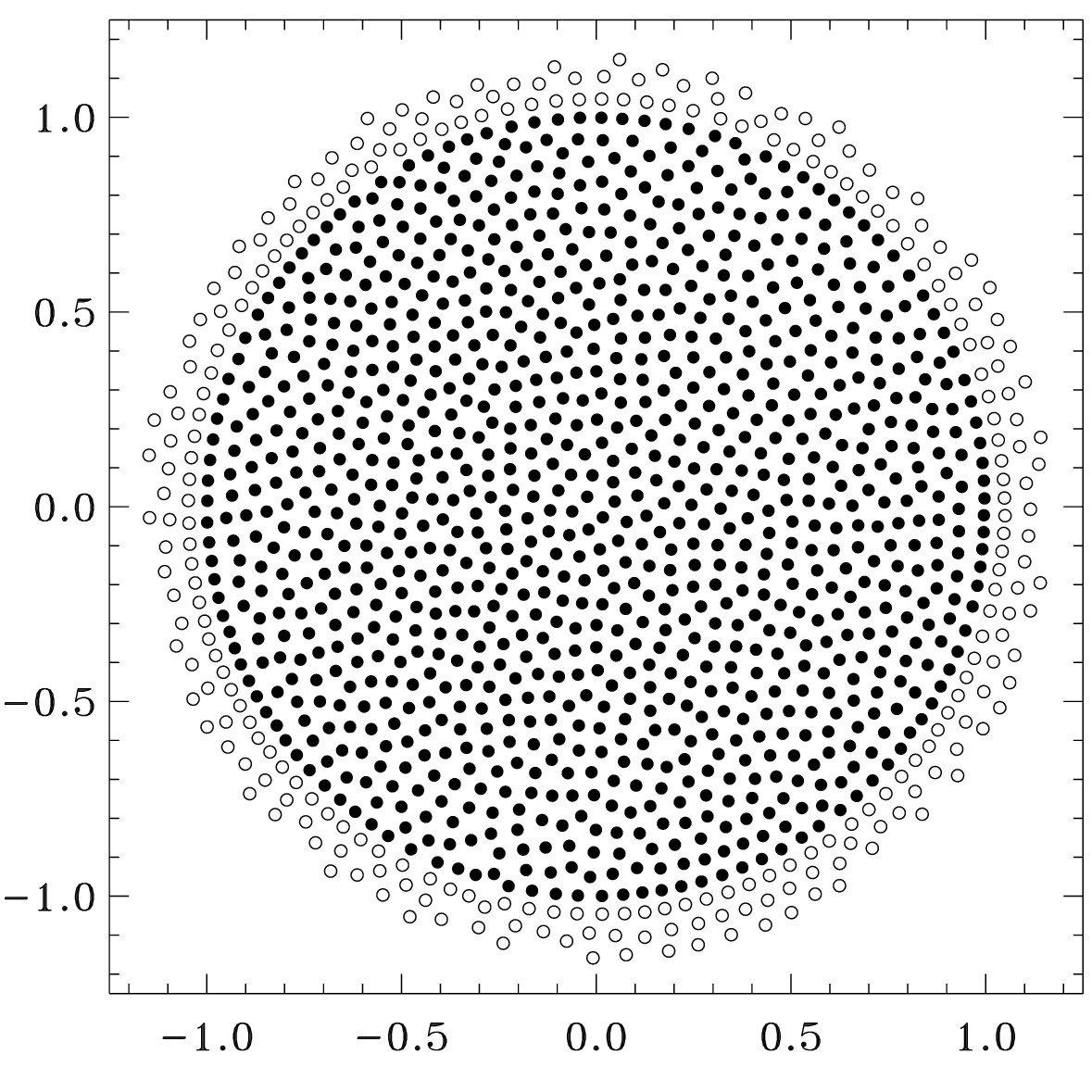}
    \includegraphics[width=0.24\textwidth]{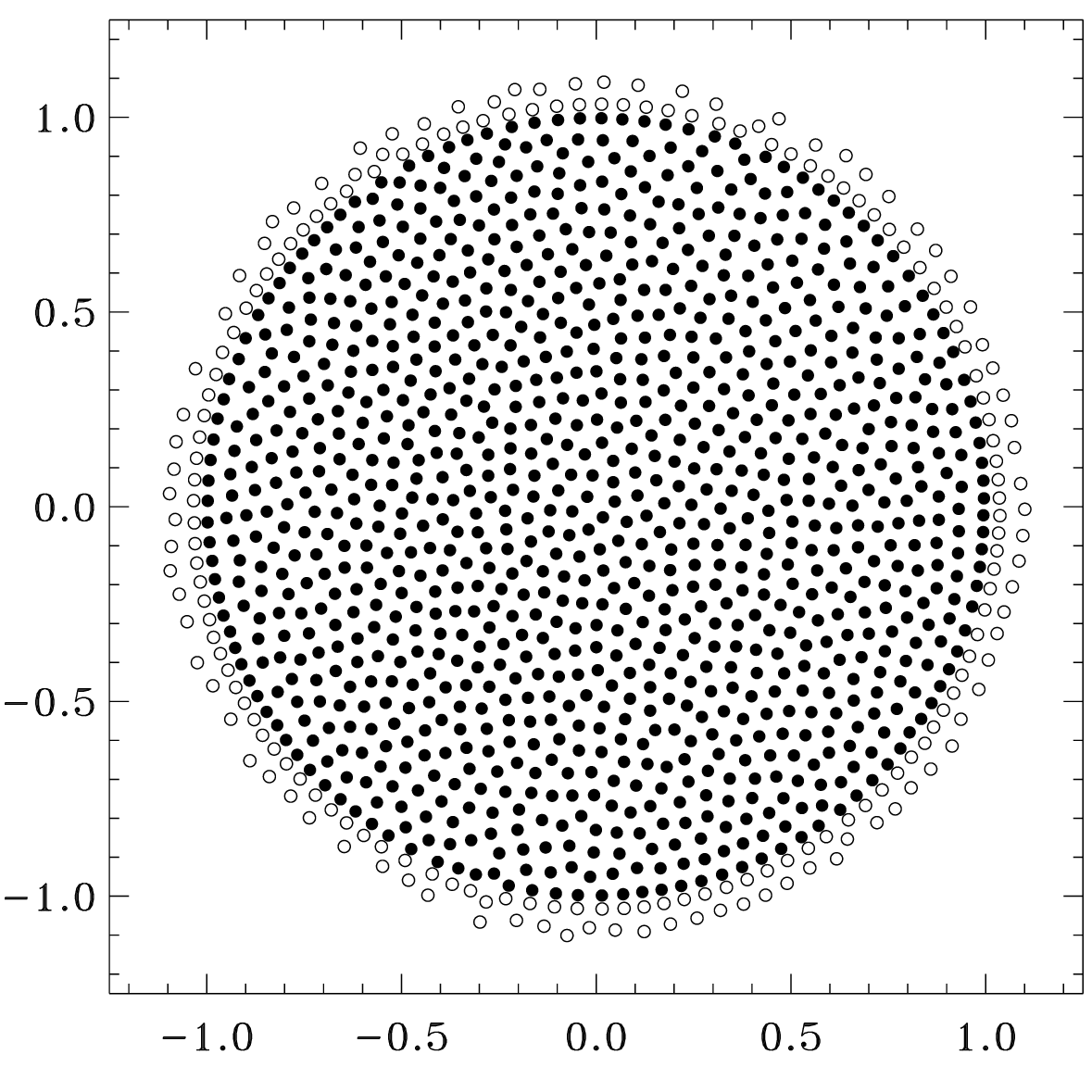}
  \end{center}
  \caption{Two-dimensional example for producing a uniform particle
    density in a circle with radius $1$ with the new WVT setup for
    1000 particles.  The frames show snapshots of the WVT iterations,
    starting with random positions sampled from a uniform distribution
    (top left), and then showing every 10th iteration, until the final
    product in the lower right panel (here, iteration 70).  The hollow
    particles are ``ghost particles'' that establish proper
    boundaries.}
  \label{f.wvtiterations}
\end{figure*}

\begin{figure*}
  \begin{center}
    \includegraphics[width=0.24\textwidth]{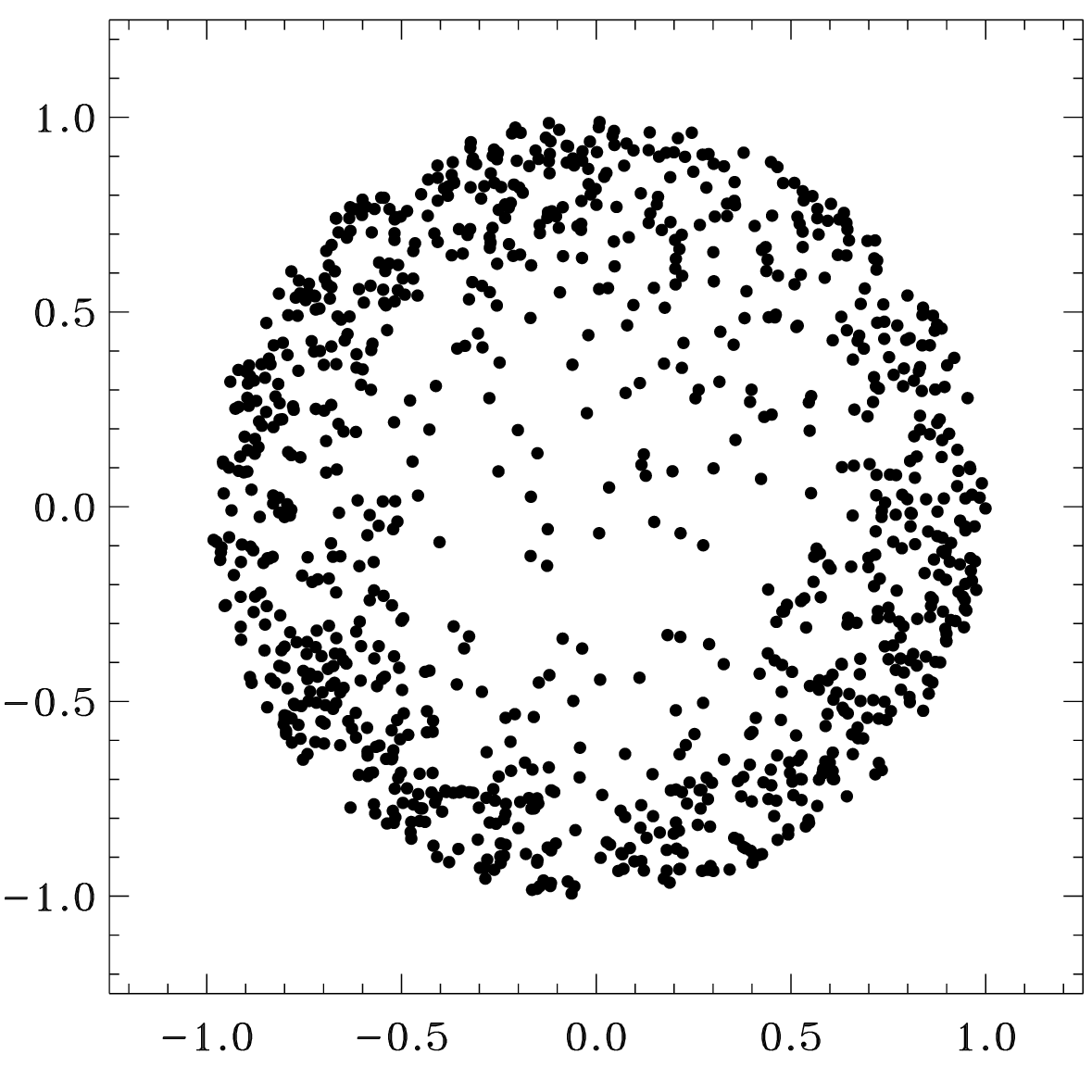}
    \includegraphics[width=0.24\textwidth]{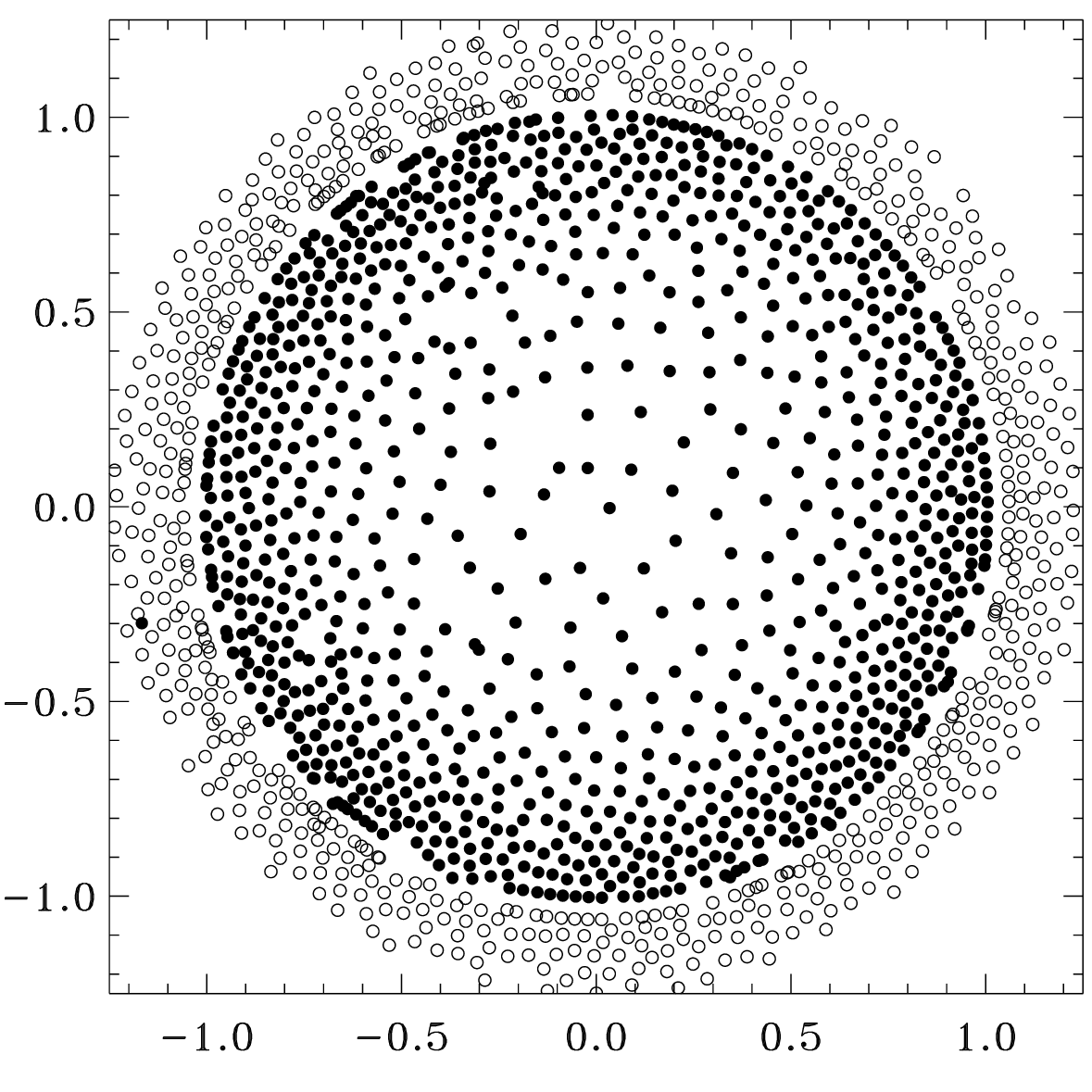}
    \includegraphics[width=0.24\textwidth]{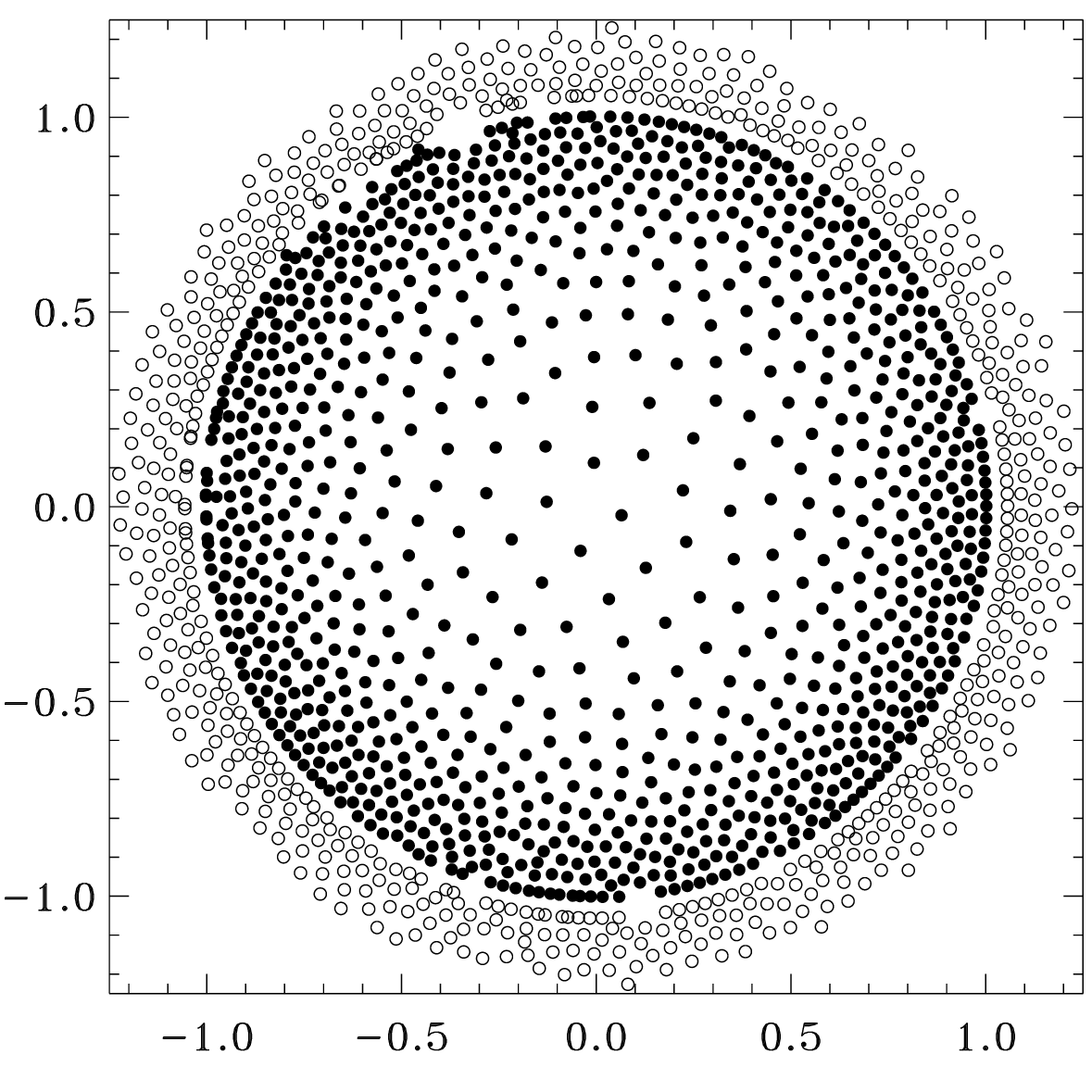}
    \includegraphics[width=0.24\textwidth]{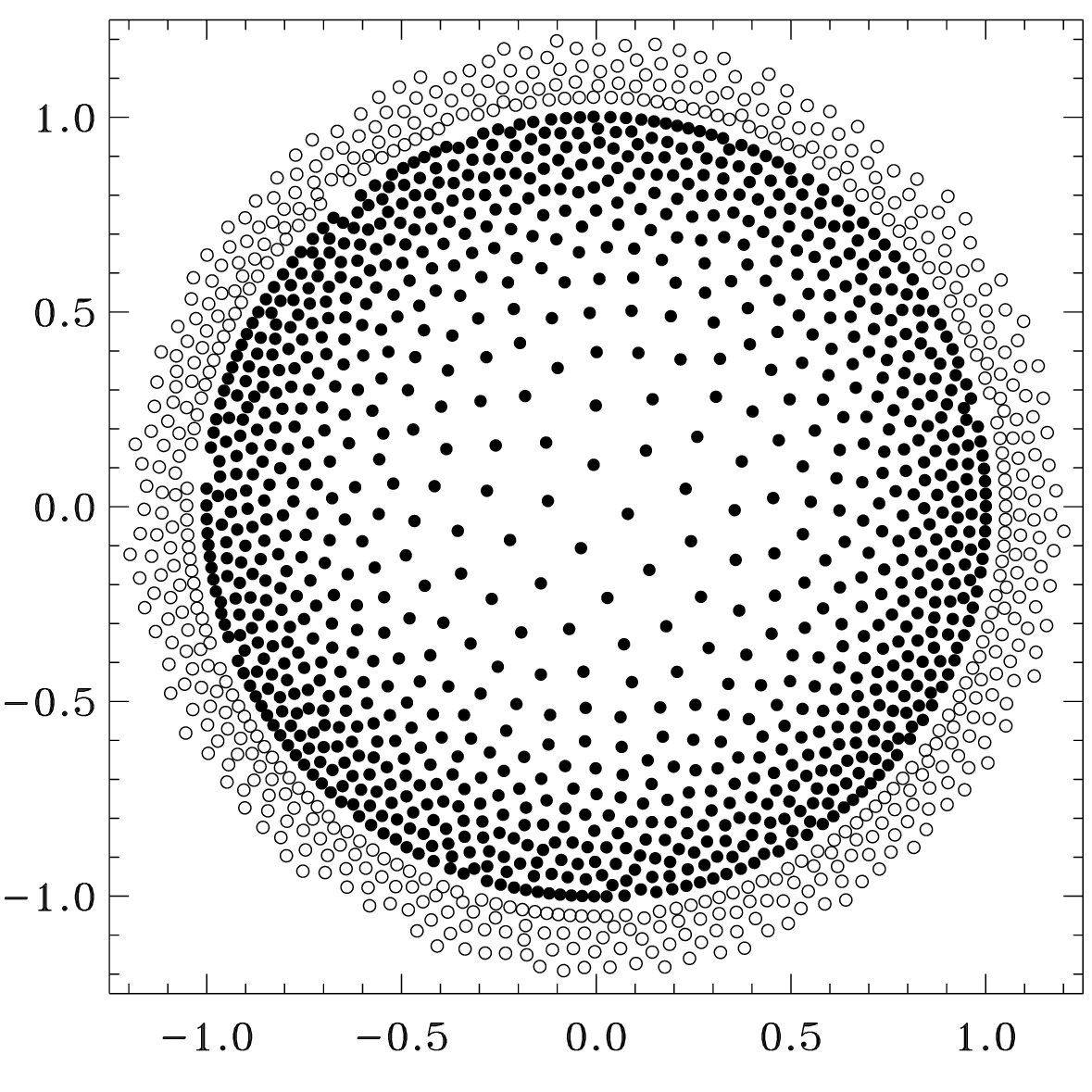}\\
    \includegraphics[width=0.24\textwidth]{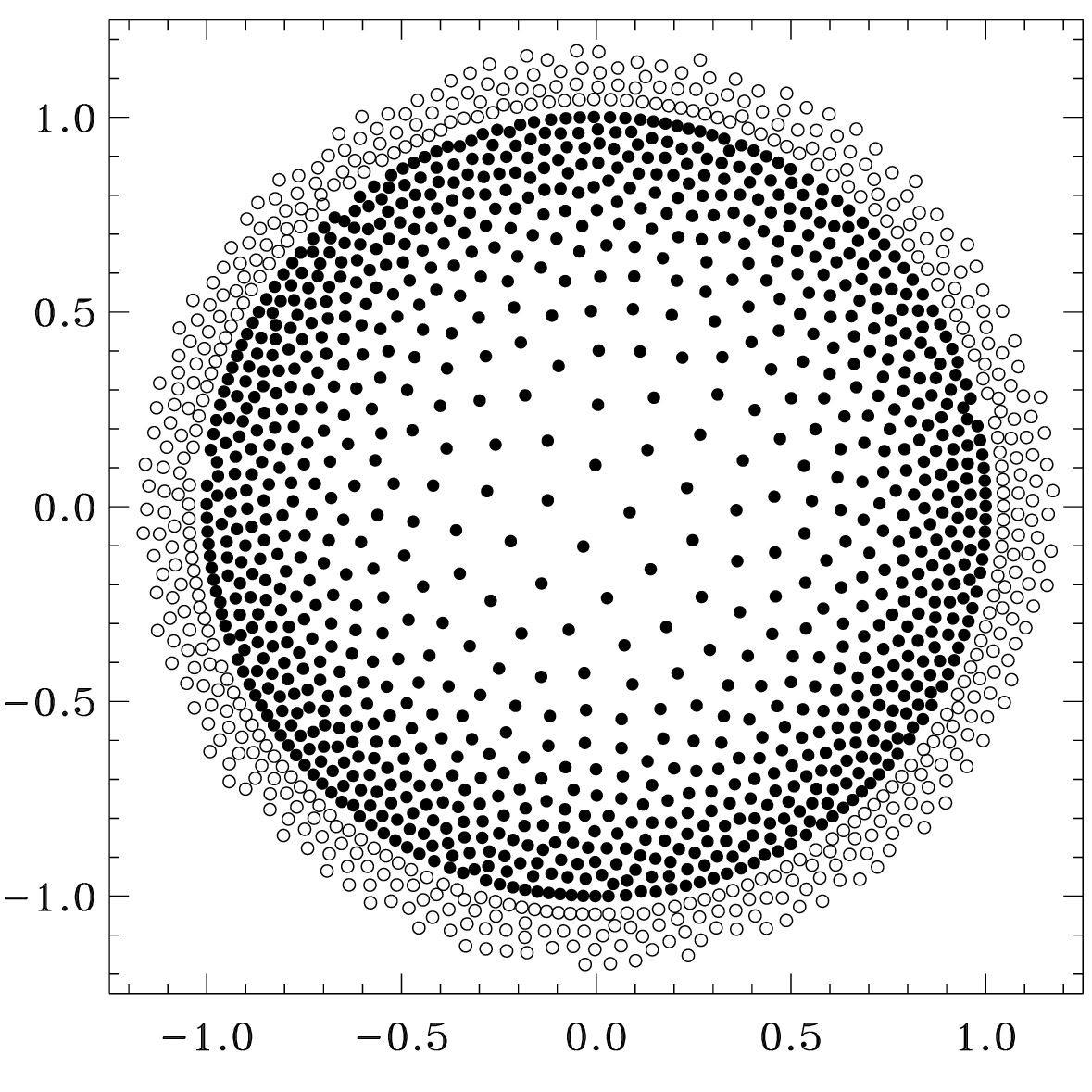}
    \includegraphics[width=0.24\textwidth]{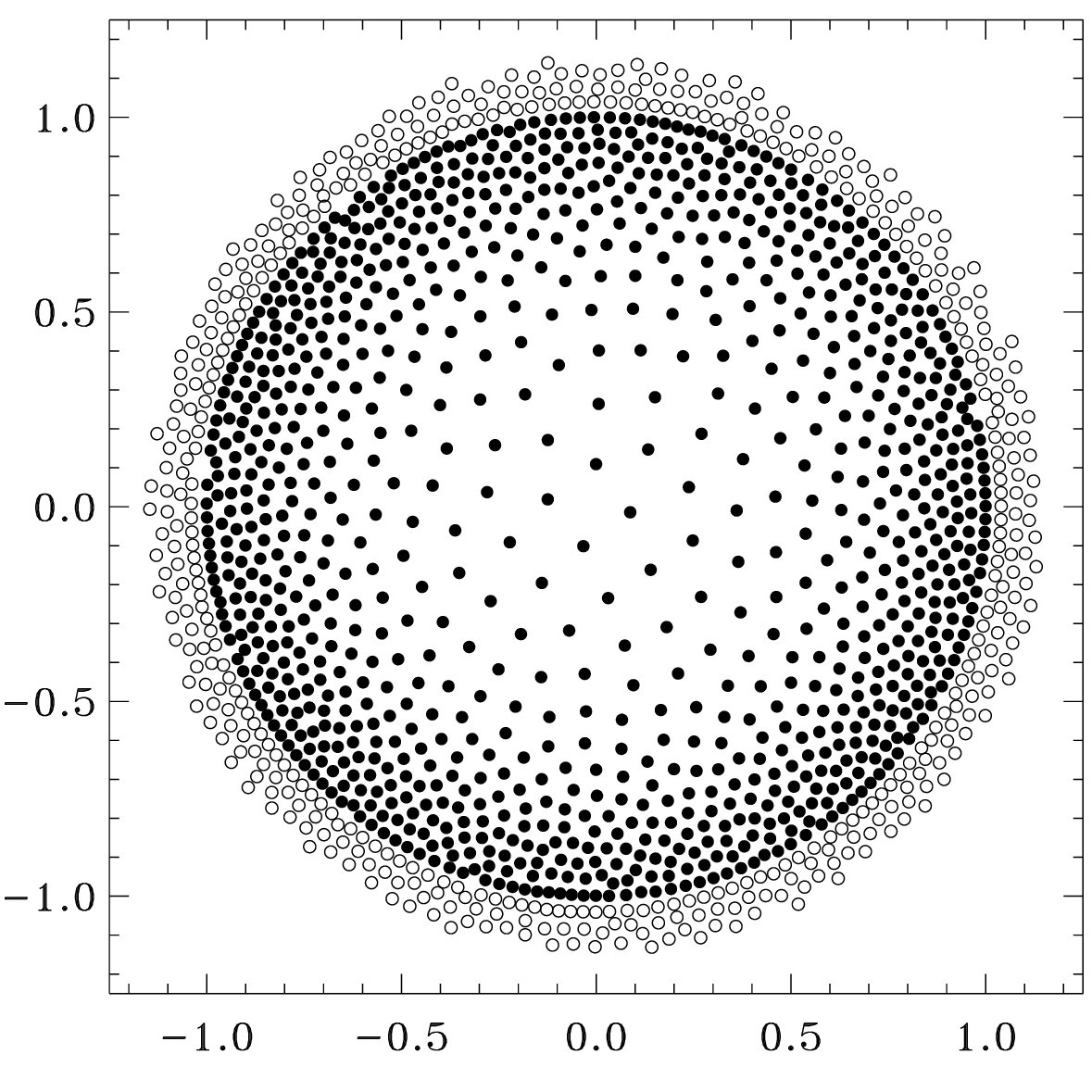}
    \includegraphics[width=0.24\textwidth]{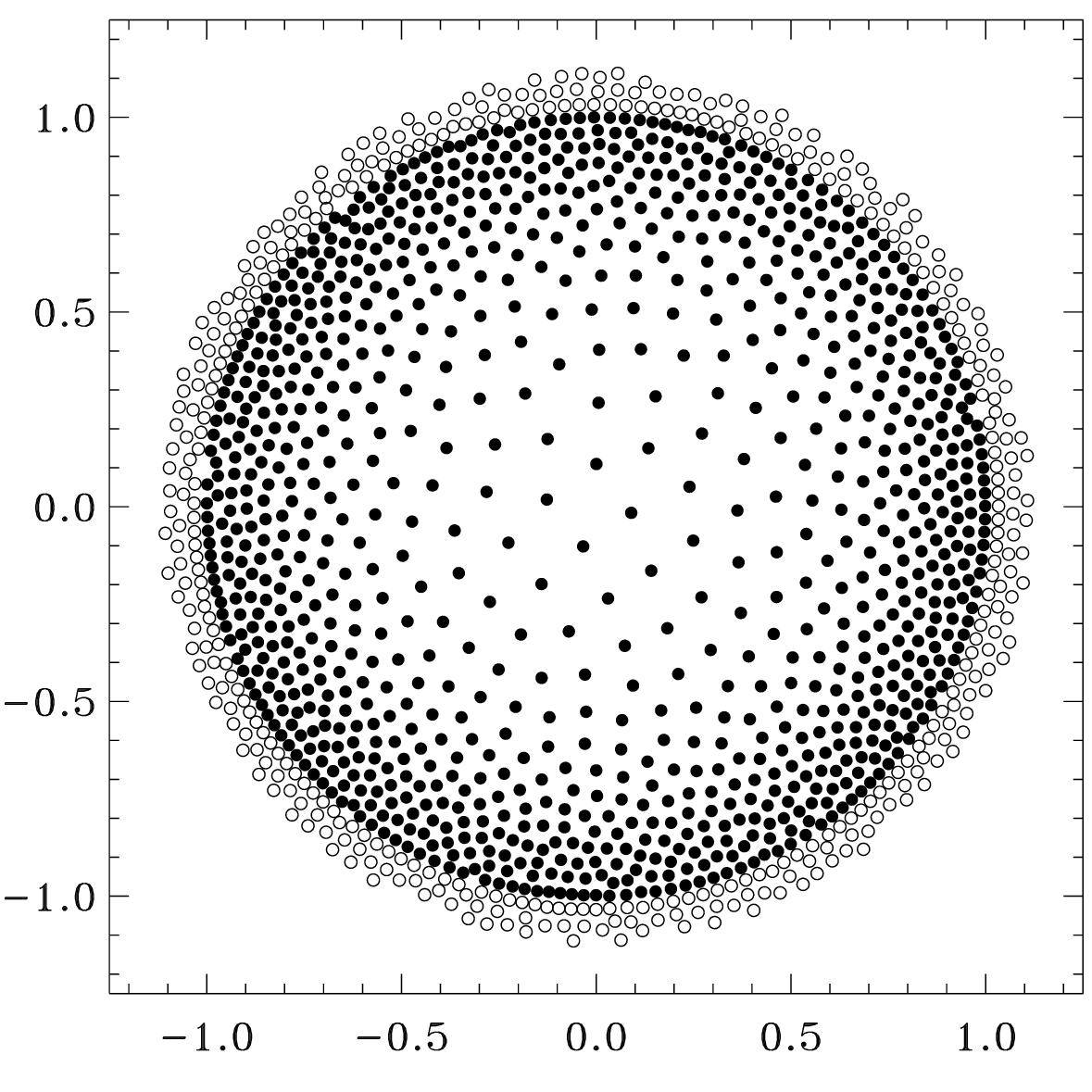}
    \includegraphics[width=0.24\textwidth]{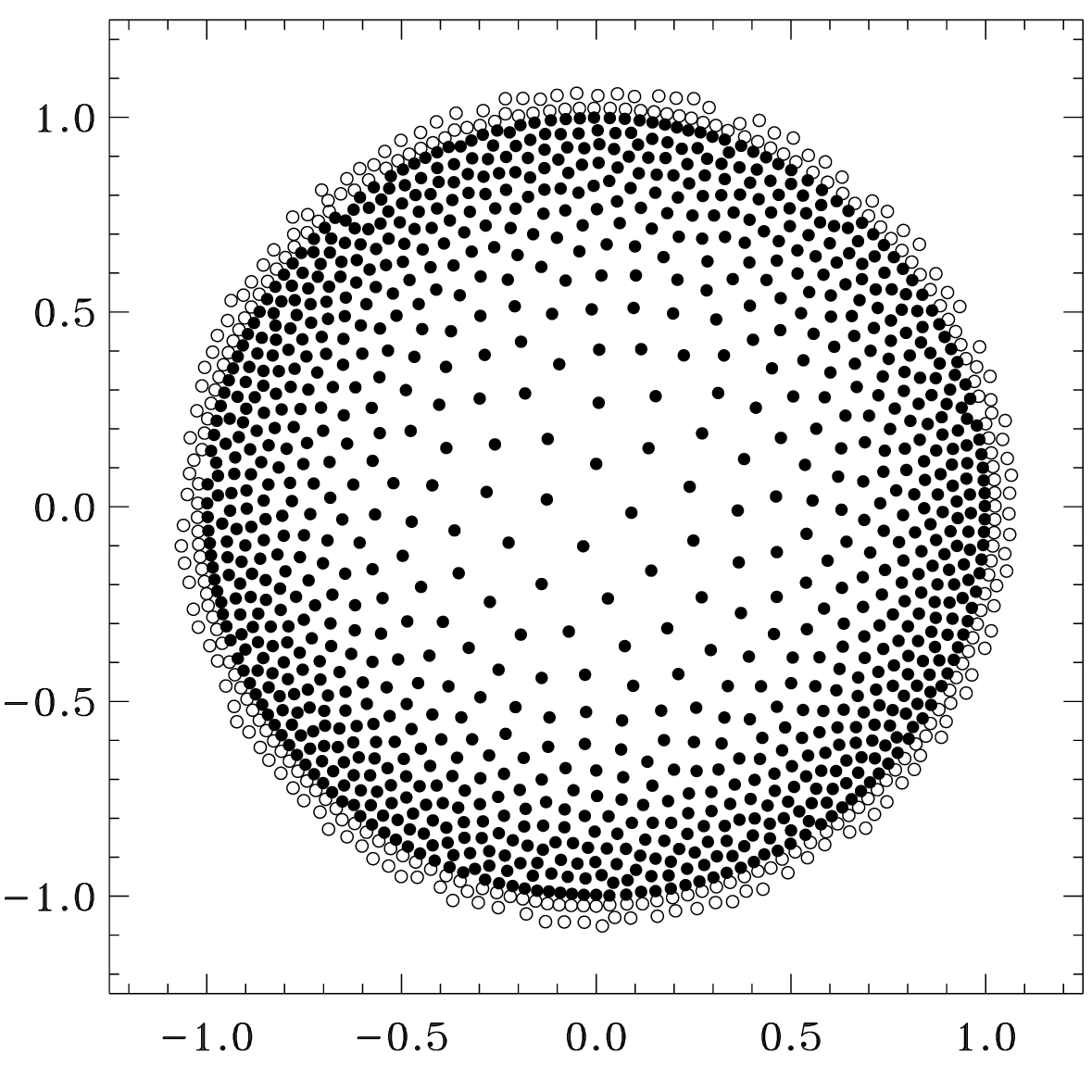}
  \end{center}
  \caption{Same as Figure~\ref{f.wvtiterations}, but for a non-uniform
    particle distribution.  The target particle density is four
    times higher at the edge of the circle than in the center.}
  \label{f.wvtnonuniform}
\end{figure*}

Given a desired average distance to neighbors $\delta({\bf r})$ at
each point ${\bf r}$ in the problem, the corresponding smoothing
length $h$ is determined by $h({\bf r})=N_{\rm neigh}^{1/n}\,
(\delta({\bf r})/2)$ in $n$ dimensions, where $N_{\rm neigh}$ is the
target number of neighbors.  We construct an initial set of particles
by sampling random positions according to the underlying particle
probability distribution $P({\bf r})\propto h({\bf r})^{-3}dV$ for a
volume $dV$.  We then evolve this configuration for multiple iteration
steps by applying repulsive forces between the particles so they
settle in the desired places. Figure~\ref{f.wvtiterations} shows a
two-dimensional example of the whole iteration process for a uniform
distribution of particles, starting with the initial collection of
particles in the upper left and ending with the final product in the
lower right.  Figure~\ref{f.wvtnonuniform} shows an equivalent
sequence for an azimuthally symmetric but non-uniform distribution.
At the outer boundary, ghost particles exert a purely radial force on
particles inside the domain, as if a smooth surface surrounded the
domain, which leads to the formation of an unusually uniform ring of
particles within 2$h$ of the boundary.  Such smoothness at the
boundary might be an asset in some simulations---for example, if the
simulation domain really is the inside of a sphere---but in other
cases the outer particles simply behave like the outermost layer of a
concentric shell configuration (\S\ref{s.spatially_adaptive}).

To reach a useful final configuration, these artificial forces should
satisfy two main requirements:
\begin{enumerate}
\item The net ``force'' on any particle should be zero (to a
  reasonable level of precision) in the desired equilibrium
  configuration, i.e., when the distance between two particles is
  identical (or very close) to the desired resolution.
\item The ``forces'' should be such that the net displacement from the
  last position is on the order of a fraction of the desired
  resolution. This will ensure equally fast convergence for high and
  low-resolution regions of the problem.
\end{enumerate}

Thus, Equation (\ref{e.wvt}) dictates the value of the repulsive
``displacement'' from bin $j$ on bin $i$, which we express for
simplicity directly as a net displacement
\begin{equation}
  \label{e.wvtdisplacement}
  \Delta{\bf x}_{i}=\sum\left[\mu\, h_{i}\,f(h_{ij},r_{ij})\,\hat{\bf r}_{ij}\right] = \mu\, h_{i} \sum\left[\,f(h_{ij},r_{ij})\,\hat{\bf r}_{ij}\right], 
\end{equation}
with $h_{ij}=(h_i+h_j)/2$. The function $f(h_{ij},r_{ij})$ should be
compact within $2h$, and empirical tests show fastest convergence of
the method for a $r^{-2}$ dependence. In practice, we add an
$\epsilon$ term in the denominator to avoid numerical problems for
close particles, and subtract a constant value to make sure the
function value vanishes at the boundary at 2$h$ and is set to 0 if the
separation is larger than that. Thus, we can express the function as
$f(h_{ij},r_{ij})=[h_{ij}/(r_{ij}+\epsilon)]^{2}+{\rm const}$. The
value of $\mu$ in Equation (\ref{e.wvtdisplacement}) regulates what
fraction of $h_{ij}$ the particles are allowed to move during each
iteration step. This free parameter should be chosen to ensure fast
convergence. In practice, we shrink $\mu$ monotonically with the
number of iterations, so particles can move relatively freely at the
beginning and at the end ``freeze'' into their final position. Note
that the particle spacing $\delta({\bf r})$ can be an almost arbitrary
function of space, as long as its value does not change significantly
across one particle spacing to ensure convergence of the method,
i.e. $\Delta\delta({\bf r})/\delta({\bf r}) <<1$.

We chose the $r^{-2}$ dependence as it reproduces locally many
desirable properties of a gravitational glass.  A functional form
based on the SPH kernel would be another natural choice for this
problem, though we did not thoroughly test this possibility.

\subsection{Practical Implementation}

Here we provide some practical advice on implementing the WVT setup on
top of an existing SPH code.

\paragraph*{Normalizing $h({\bf r})$:}

In our description of the WVT method above, we have assumed that we
know \textit{a priori} the desired particle spacing $\delta({\bf r})$
as a function of position. However, it can be difficult to guess what
specific spatial resolution, at each point in a complex problem, will
produce a configuration with an acceptable total particle count. Thus,
our implementation interprets the input particle spacing as a relative
rather than an absolute value, and scales it according to a desired
number of particles $N_{\rm SPH}$ and neighbors $N_{\rm neigh}$. At
each iteration step, $h({\bf r})$ is evaluated for all particle
positions and we then compute the sum of all individual SPH particle
volumes: $V_{\rm SPH}=\sum_i[ (4\pi/3)(2\,h_i)^3]$. Since we do know
the actual volume $V$ of our computational domain and that we desire
$N_{\rm neigh}$ neighbors within 2$h$ for each particle, we scale all
$h({\bf r})$ values so that $V=V_{\rm SPH}/N_{\rm neigh}$.

\paragraph*{Treating boundaries:}

Most modern SPH codes have some kind of boundary treatment already
implemented. Fixed boundaries are usually implemented by means of
ghost particles \citep[e.g.,][]{HerantDirtyTricks} that exert
antisymmetric forces on the particles to keep them across a given
boundary. WVT works well with this type of boundary treatment, and we
suggest mirroring SPH particle layers within 2 smoothing lengths at
the boundary interface and adding them to the set of normal particles
during the pseudo force calculation step. Periodic boundaries may also
be used, and the region of interest can simply be ``cut out''
afterwards. We find this method to work well for arbitrary geometries.

\paragraph*{Updating particle positions:}

The most convenient way to implement WVT is to use the entire
structure of your existing SPH code with as few changes as
possible. We suggest modifying the existing SPH loop to calculate the
sum in Equation (\ref{e.wvtdisplacement}) and then multiply this
pseudo ``velocity'' by the ``individual time steps'' $\mu h_i$ to
update particle positions. If the distribution does not appear to
converge, we suggest decreasing the value of $\mu$ with each
iteration.

\paragraph*{Finishing the iterative process:}

Judging when an initial setup is sufficiently good is
application-dependent and at least somewhat subjective.  In our
experience, slowly reducing the value of $\mu$ (the maximum fractional
distance a particle can be moved in one time step; see Equation
\ref{e.wvtdisplacement}) works well.  For the cases we've studied,
which include setups using from 100,000 to 50 million particles
\citep{RaskinWD1,RaskinWD2,FryerWDmerger,Passy11,Ellinger11},
convergence occurs in about 100 iterations, usually even after only 40
iterations. However, this number may strongly depend on the details of
the algorithm, and the desired interpolation accuracy, an issue that
we discuss in more detail in section \ref{s.comparison}.


\section{Example Applications}
\label{s.examples}

Different applications may impose very different requirements on the
resolution of a particular object.  When particle mass density and
desired number density vary together, WVT can generate initial
conditions with particles of uniform mass; when the mass density and
desired number density vary in different ways, WVT can produce
configurations where both particle mass and size vary across the
domain.  We now consider different examples, mostly from stellar
interactions, that impose very different numerical requirements.

\subsection{Uniform Particle Density}

In the first application, we consider representing a star with a
uniform particle density, which would be appropriate for simple
head-on collisions between two stars. In this situation, it is very
likely that both center and outer layers will be heavily involved in
the process, as all parts should be equally affected during the
merger, and would require equal resolution to resolve hydrodynamic
effects throughout the star.

\begin{figure*}
  \begin{center}
    \includegraphics[width=0.3\textwidth]{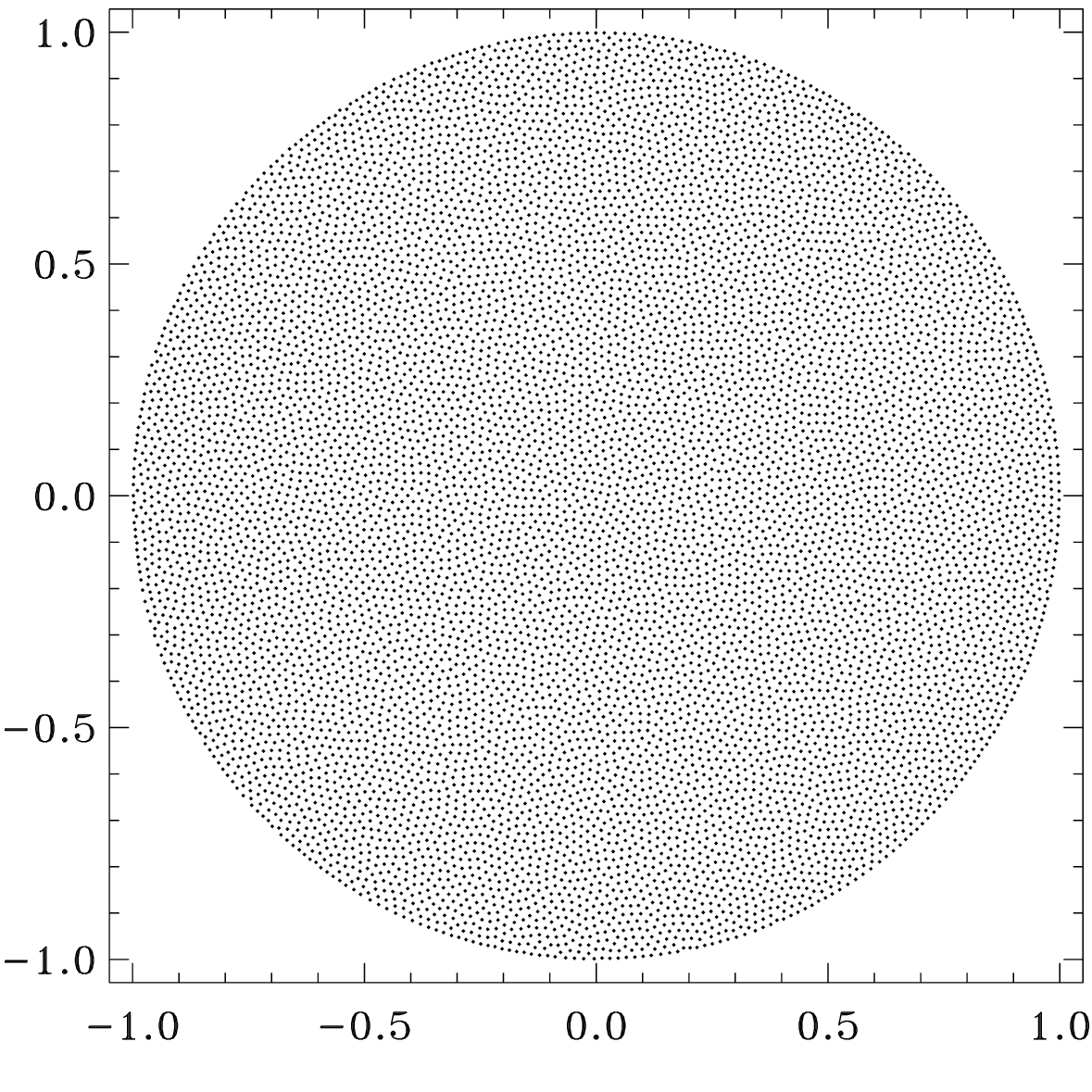}
    \includegraphics[width=0.3\textwidth]{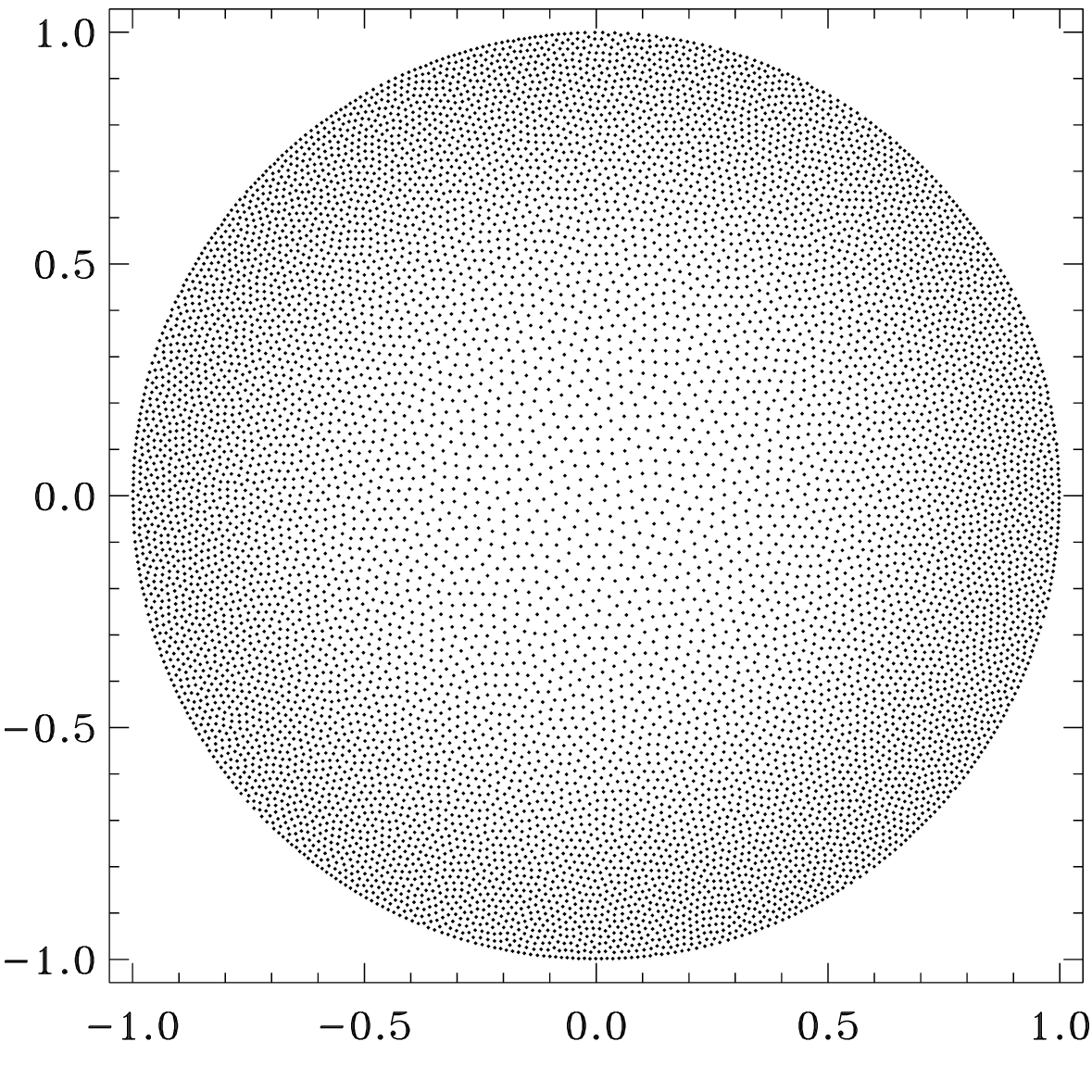}
    \includegraphics[width=0.3\textwidth]{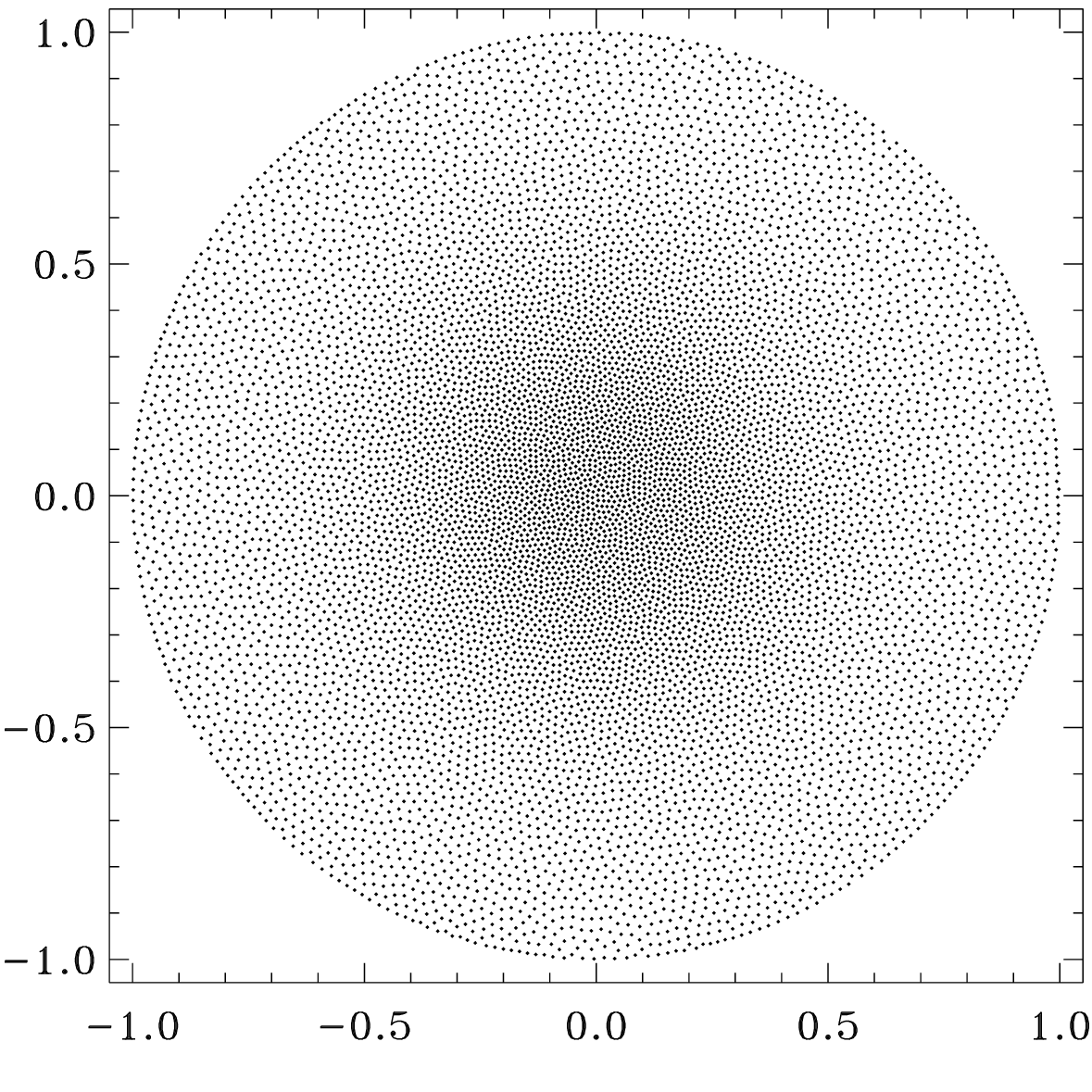}\\
    \includegraphics[width=0.3\textwidth]{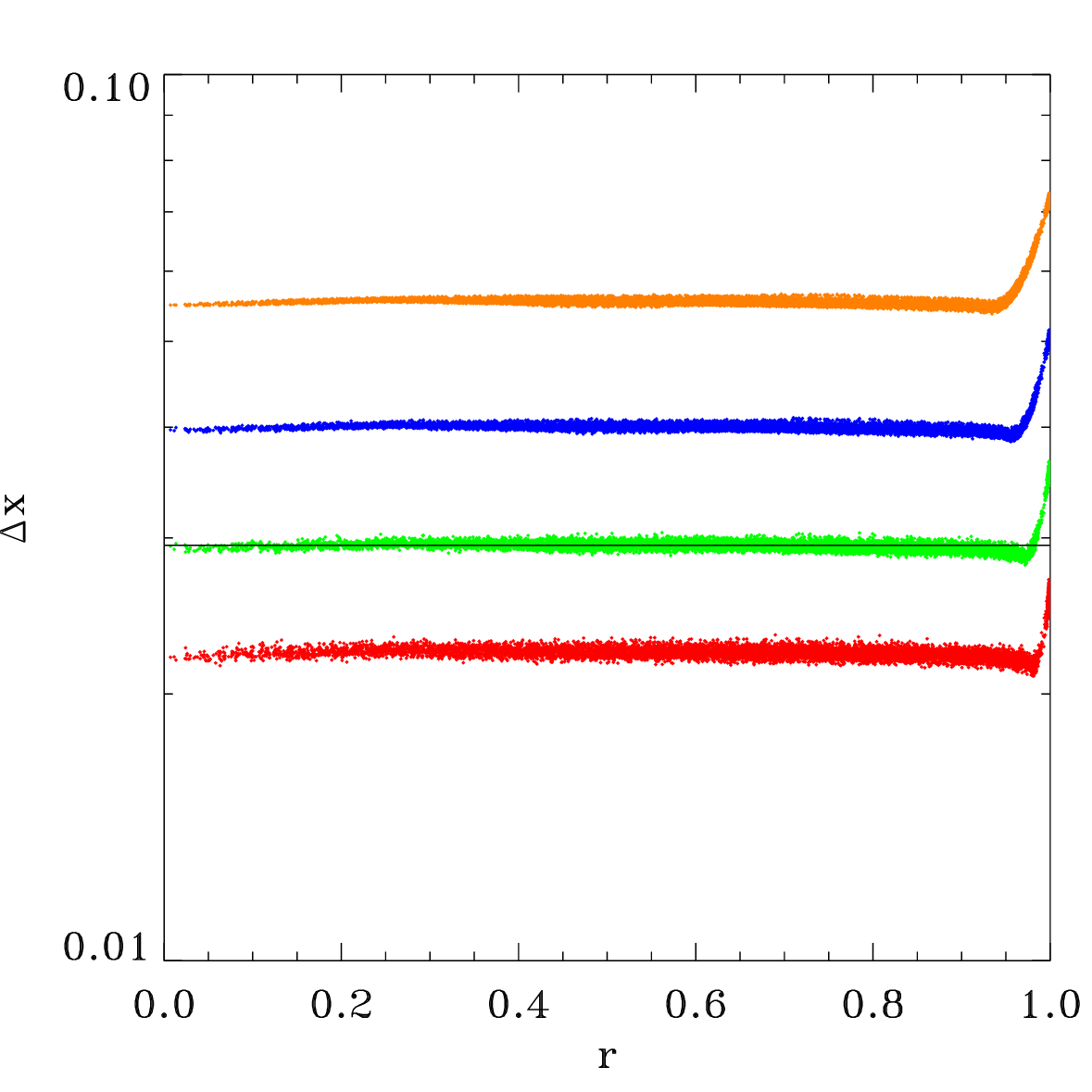}
    \includegraphics[width=0.3\textwidth]{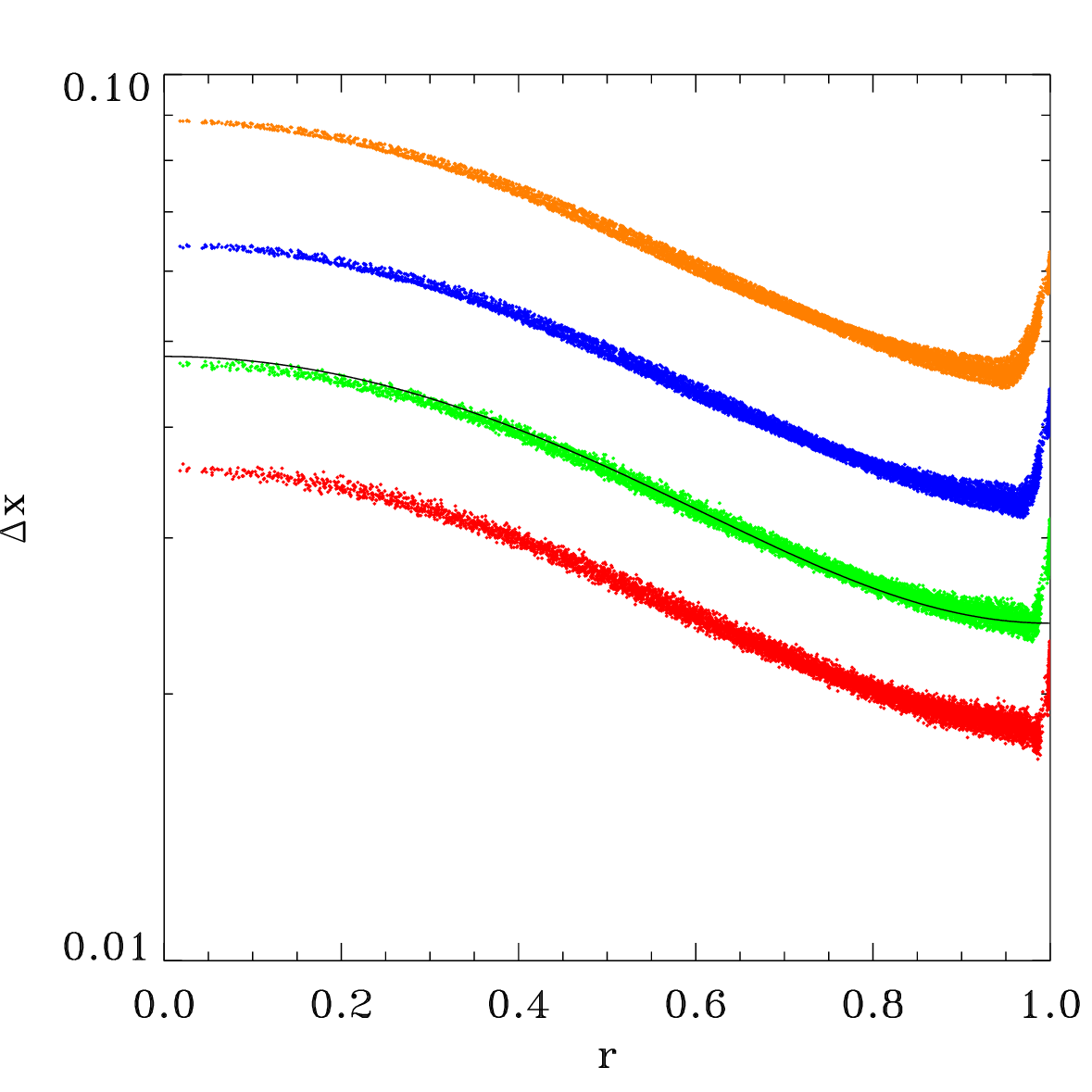}
    \includegraphics[width=0.3\textwidth]{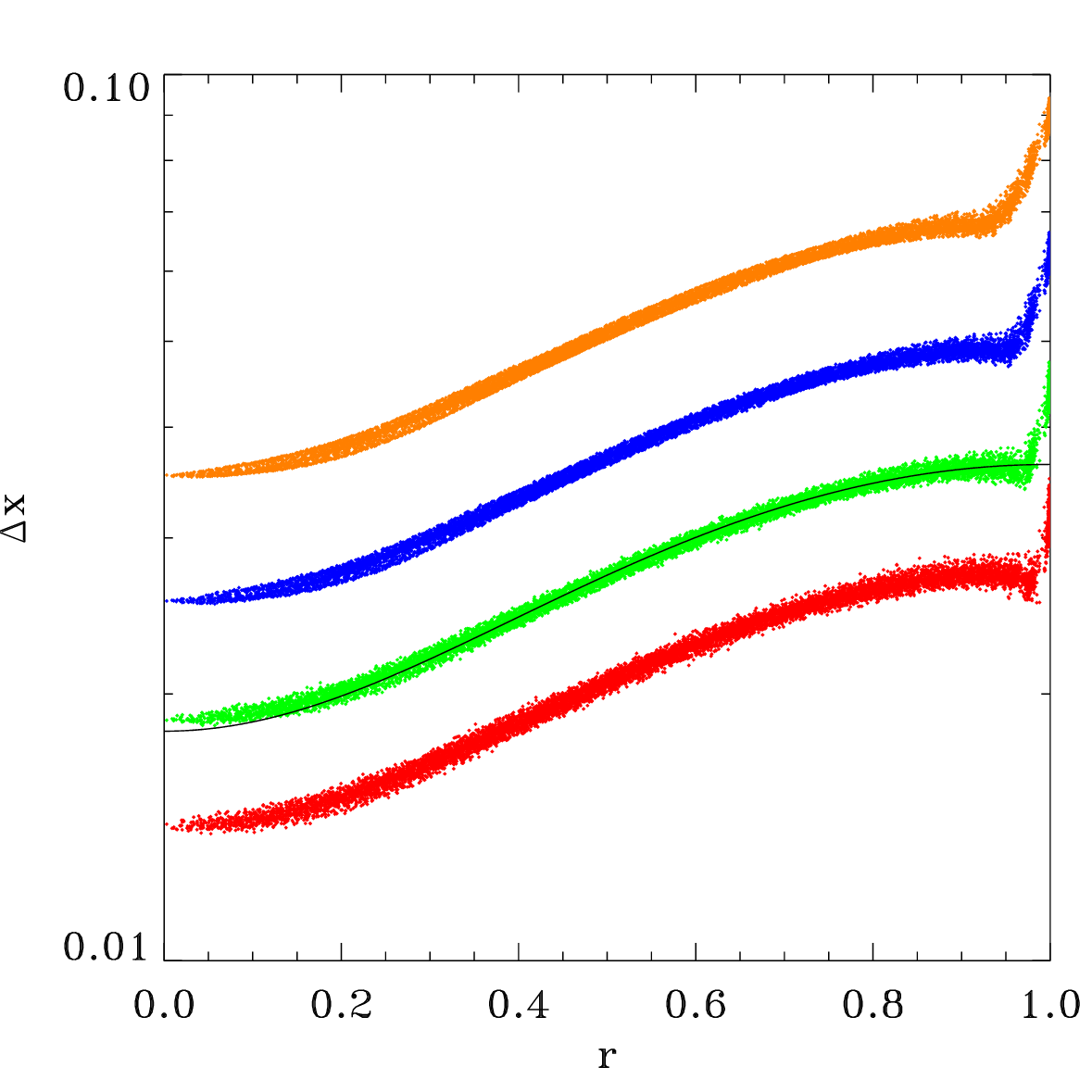}
  \end{center}
  \caption{Top panels: particle configurations in two-dimensional
    examples. We consider three different configurations: uniform
    particle density (left), more resolution in outer layers (center),
    more resolution at center (right). The bottom panels shows the
    actual particle separations as a function of radius. The points
    measure the average distance to the closest 8 (red), 16 (green),
    32 (blue) and 64 (orange) neighbors. The solid black line shows
    the input resolution scaled for the closest 16 neighbors, indeed
    closely following the green data points.}
  \label{f.polytrope}
\end{figure*}

Such a 2D particle setup is shown in the top left panel of
Figure~\ref{f.polytrope}. The lower panel shows how well the desired
resolution is achieved by measuring the average distance between
particles for the closest 8 (red), 16 (green), 32 (blue) and 64
(orange) neighbors. The black solid line shows the expected
theoretical values for 16 neighbors. The deviations at the boundary
are due to the lack of neighbors across the boundary, and can of
course be fixed by increasing h accordingly in that region.

\subsection{More Resolution in the Center}

However, if the user is interested in any type of mixing within the
stars or during a merger, it is imperative to use equal-mass
particles. Work by \citet{LombardiMixing} suggests that artificial
forces between non-equal mass particles lead to numerical diffusion
and artificial mixing. With the WVT setup, we can enforce equal mass
particles by adjusting the particle separation according to the
underlying density $\rho({\bf r})$, such that $\delta({\bf r})\propto
\rho({\bf r})^{-1/3}$. This results in a setup with more resolution in
the center, as shown in the right panel of Figure~\ref{f.polytrope}.

\subsection{More Resolution in the Outer Layers}

If one is interested in studying the more gentle Roche Lobe overflow
phase in a binary, one needs as much numerical resolution as possible
in the outer layers of the donor star, in order to sufficiently
resolve the overflow and accretion stream. The middle panel of
Figure~\ref{f.polytrope} shows a polytrope where we have put more
resolution in the outermost layer than in the center.

\subsection{Asymmetric Initial Conditions: Double Degenerate Binary}

\begin{figure*}
  \begin{center}
    \includegraphics[width=0.75\textwidth]{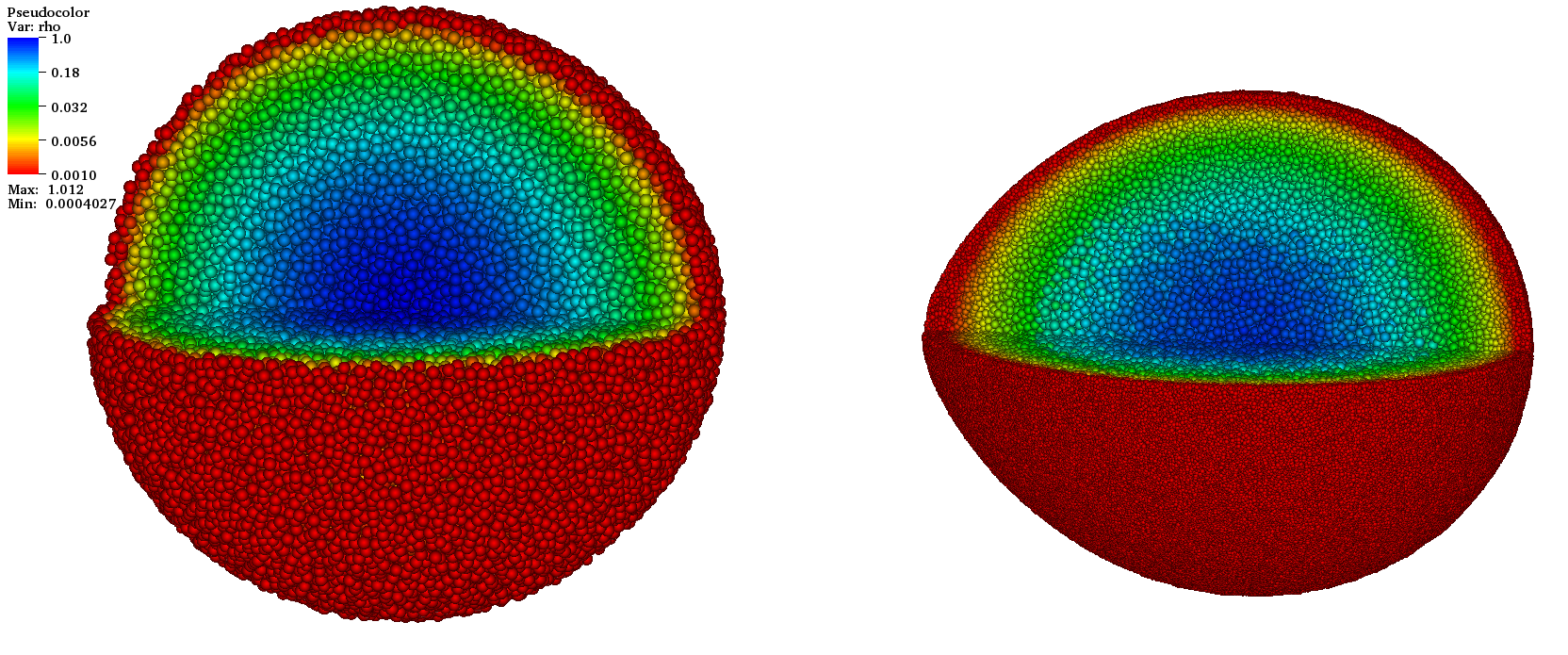}
  \end{center}
  \caption{This three-dimensional example shows an asymmetric WVT
    setup for a double degenerate merger simulation. In this example,
    the accretor (left) is modeled with a constant particle density,
    whereas the donor (right) has significant more resolution in the
    outer layers than in the center, making SPH simulation of Roche
    lobe overflow feasible.}
  \label{f.dd}
\end{figure*}

Figure~\ref{f.dd} shows an example of an asymmetric, three-dimensional
setup with WVT. The picture depicts a double degenerate binary system
with the donor (right) on the verge of overflowing its Roche
lobe. Note how the size of the SPH particles (varying sizes and colors
of spheres) is much smaller in the outer layers of the donor, which
helps significantly in resolving the mass transfer stream in the
simulation \citep{DiehlDDmerger}.  The evolution of the system is
extremely sensitive to the initial mass transfer (as it governs the
evolution of the orbit).  Resolving this mass transfer is critical to
achieving good agreement between SPH and rotating grid simulations.

\subsection{Asymmetric Initial Conditions: Elliptical Galaxies}

Simulations of normal elliptical galaxies commonly use a gas component
embedded within a dark matter halo in order to model evolution
properly.  Creating an initial configuration for a simulation of
feedback effects in such a galaxy, it is reasonable to say the gas has
settled hydrostatically into the dark matter potential and is
quiescent.  We assume that the gravitational effects of the gas are
small compared to those of the dark matter.  Then, we should expect
the gas density $\rho_{\textnormal{gas}}$ to follow the dark matter
potential $\Phi_{\textnormal{DM}}$, assuming a polytropic equation of
state:

\begin{equation}
  \label{polytrope}
  \rho_{\textnormal{gas}}(x,y,z) = \left[ \frac{-1}{K (n+1)} \Phi_{\textnormal{DM}}(x,y,z) + C \right] ^{n}
\end{equation}
where $K$ is a constant, $n$ is the polytropic index, and $C$ is a
constant of integration that determines the sharpness of the edge of
the density distribution.  Note that $\rho_{\textnormal{gas}}$ and
$\Phi_{\textnormal{DM}}$ need not be spherical in shape; the dark
matter potential maps into the gas density regardless of its degree of
eccentricity.

In fact, we may take direct advantage of the non-requirement for
spherical shape; the WVT code contains an option to use a cloud of
dark matter particles as three-dimensional interpolation points in
order to determine the value of the dark matter potential at a given
position.  Then, Equation \ref{polytrope} defines the mass density for
a gas particle placed at that point.

\begin{figure*}
  \begin{center}
    \includegraphics[width=\textwidth]{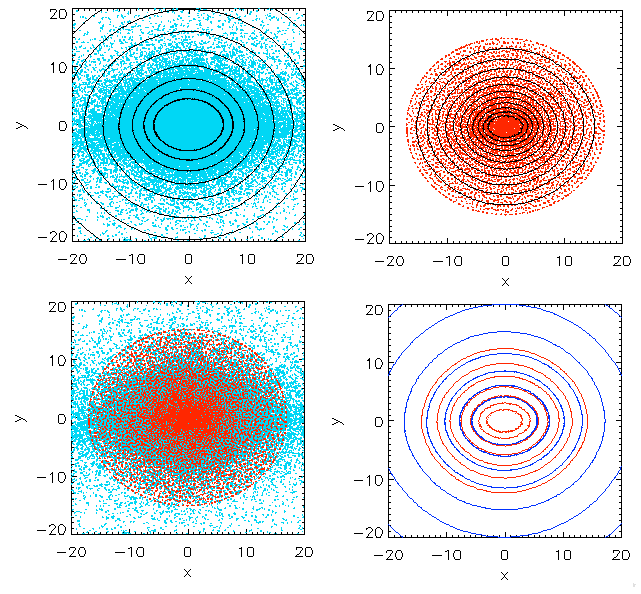}
  \end{center}
  \caption{WVT results for a gas cloud with polytropic index of $n =
    3/8$, embedded within a triaxial (axis ratio 17:15:14) dark matter
    potential, a slice through the simulation at $z = 0$.  Top-left:
    Dark matter particles in cyan, with surfaces of constant potential
    overplotted.  Top-right: SPH gas particles in red, with surfaces
    of constant mass density overplotted.  Bottom-left: SPH gas
    particles in red and dark matter particles in cyan.  Bottom-right:
    The surfaces of constant dark matter potential (blue) coincide
    with the surfaces of constant gas mass density (red).}
  \label{ellip_contours}
\end{figure*}

If Equation \ref{polytrope} is applied correctly, the resulting
surfaces of constant gas mass density must coincide with the surfaces
of constant dark matter potential.  Thus, as a diagnostic test, we
first construct a self-consistent Hernquist sphere of N-body particles
with a distribution function of Ossipkov-Merritt form
\citet{1979SvAL....5...42O, 1985MNRAS.214P..25M, 1985AJ.....90.1027M,
  2004ApJ...601...37K} and anisotropy radius $r_{a} = 1 \times
10^{10}$, and then use the method of \citet{2001ApJ...549..862H} and
\citet{2008ApJ...679.1232W} to deform the sphere adiabatically into a
triaxial system.  The resulting configuration of N-body particles has
approximate axis ratio 17:15:14, and forms the set of interpolation
points from which to create a cloud of gas particles via WVT.  The
example gas cloud analyzed in Figure~\ref{ellip_contours} has a
polytropic index of $n = 3/8$, and shows gas isodensities coincident
with the dark matter isopotentials.

\subsection{Asymmetric Initial Conditions: WVT Logo}

\begin{figure*}
  \begin{center}
    \includegraphics[width=0.75\textwidth]{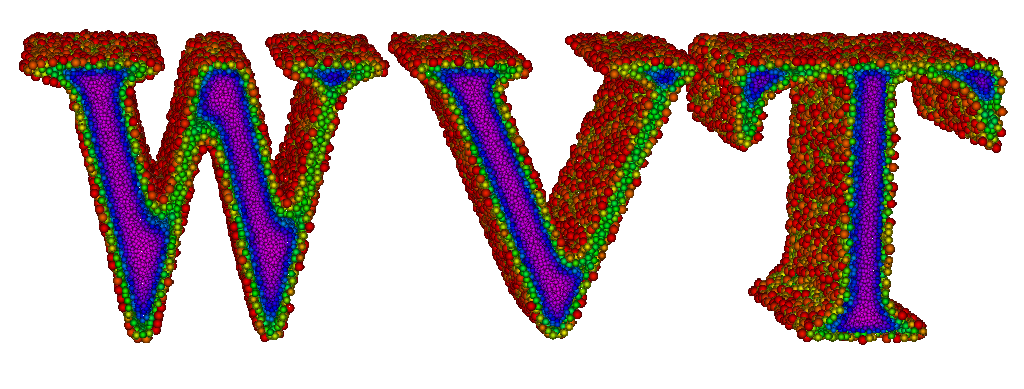}
    \includegraphics[width=0.75\textwidth]{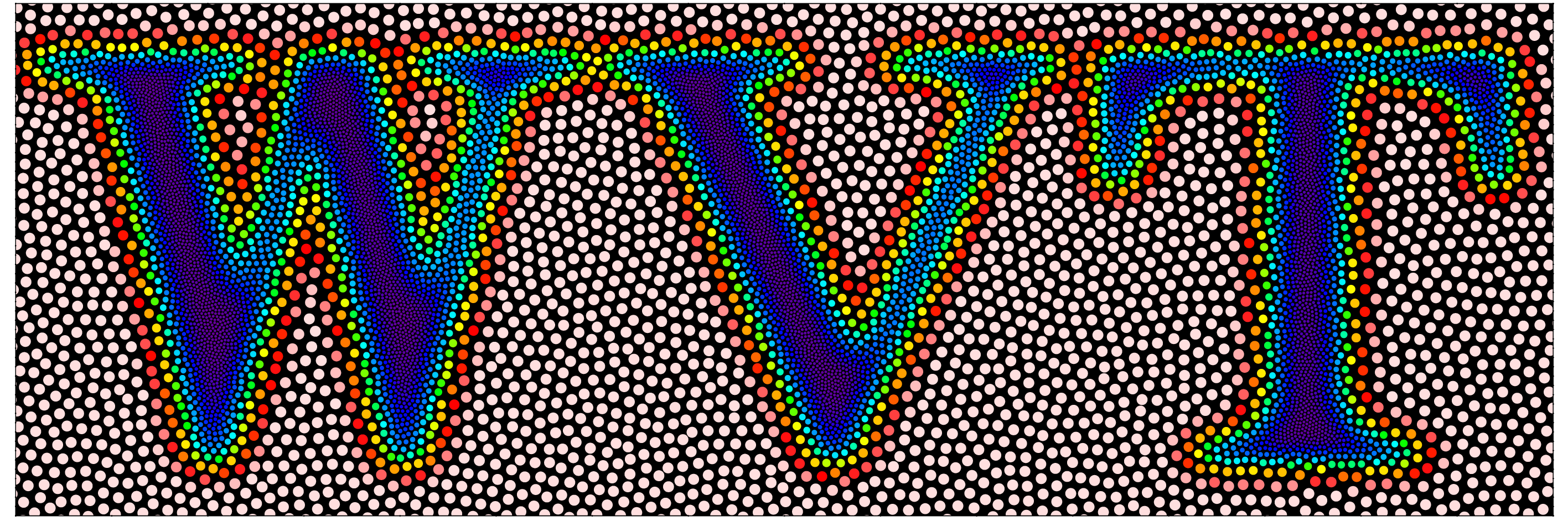}
  \end{center}
  \caption{Examples for an arbitrary spatial configuration. The top
    panel shows a 3d configuration, the lower panel shows a
    two-dimensional configuration. Particles with large smoothing
    lengths (shown in white in bottom panel) are omitted in the
    three-dimensional view for clarity. This particular configuration
    shows a dynamic range of $\sim10$. Smoothing lengths are indicated
    by color and proportional to their symbols' sizes.}
  \label{f.wvtlogo}
\end{figure*}

Figure~\ref{f.wvtlogo} shows another example of an arbitrarily complex
setup. The top panel shows the letters ``WVT'' used in three
dimensions, the lower panel gives the same example in two
dimensions. Note that we omitted the largest SPH particles (white
particles in the 2D version) in the 3D version for clarity. WVT has no
difficulties in matching the desired resolution even in these complex
test cases.

\subsection{Mixing in Supernovae}

As the shock moves out through a star in a supernova explosion,
Richtmyer-Meshkov and Rayleigh-Taylor instabilities drive turbulence
behind the shock.  This turbulence mixes elements, dredging up
radioactive $^{56}$Ni and injecting hydrogen into slower moving
layers.  This mixing is observed in supernova light-curves and in the
knots in supernova remnants.  But modeling this turbulence is not
trivial; the shock radius expands by several orders of magnitude, and
both cartesian grid Eulerian and SPH codes introduce numerical
turbulence (based on noise in the initial set-up) that can
artificially produce spurious turbulence.

The Sedov blast wave is an ideal test for any code modeling these
explosions; an analytic solution exists and can be compared to
simulation results.  This test also exposes consequences of choices in
initial conditions beyond just total particle or mesh cell count.
Approximating a point explosion in a large volume benefits from higher
spatial resolution near the origin, and the shock is often unstable to
hydrodynamic instabilities (e.g., Richtmyer-Meshkov and
Rayleigh-Taylor behind the shock), so any initial density perturbation
introduced by the setup (or grid effects in an Eulerian code) can
artificially seed turbulence.

\begin{figure*}
  \begin{center}
    \includegraphics[width=0.75\textwidth]{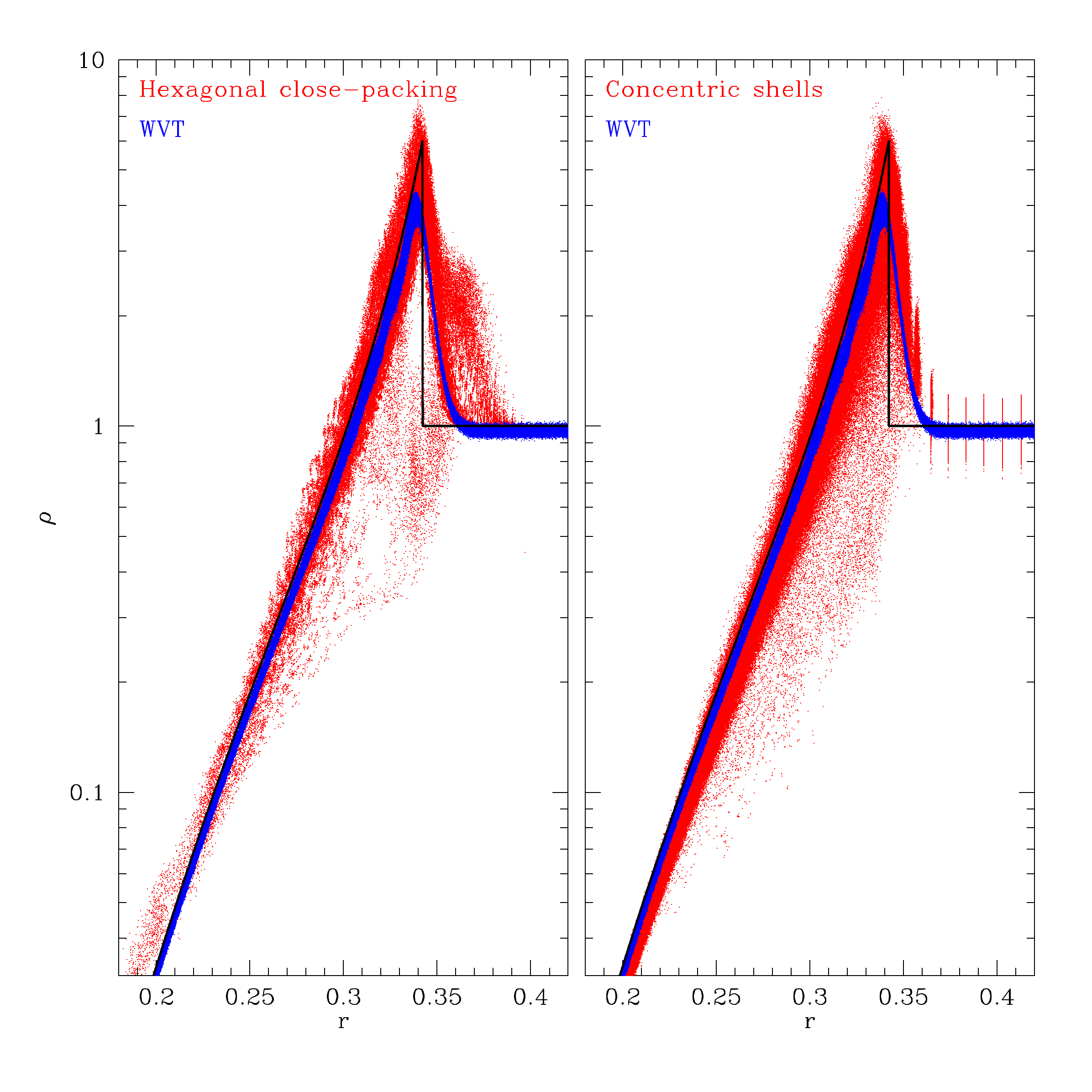}
  \end{center}
  \caption{Density versus radius of a Sedov blast wave problem
    comparing the results from a WVT setup with a hexagonal
    close-packed lattice (left) and a concentric shell configuration
    (right), each using 1.5 million particles.  The black line
    indicates the analytic solution.  In the hexagonal close-packed
    lattice, different shock velocities at different angles through
    the lattice lead to variation in the shock position around the
    sphere---and to lower-density regions behind the fastest parts of
    the shock, which show up in the plot as extra scatter in density,
    especially for radii between $\sim 0.32$ and $\sim 0.36$, at this
    time in the simulation.  The initial density perturbations in the
    concentric shell setup---visible as scatter in density at constant
    radius outside the shock---grow in the shock to produce a broad
    range of particle densities.  The low resolution at the energy
    source leads to velocity perturbations that then create density
    perturbations.}
  \label{fig:sedov}
\end{figure*}

With a given total particle budget, different schemes for generating
initial conditions have different degrees of success in meeting the
resolution, homogeneity and isotropy requirements of Sedov
simulations.  Figure~\ref{fig:sedov} shows the particle distribution,
in terms of mass density versus radius, for three different
simulations of a Sedov blast wave---one using a hexagonal close-packed
lattice, one using a concentric shell configuration, and one using
WVT---at a time $t = 0.06317$ after the launch of the shock.  The
number of particles is nearly identical for each simulation
($1.52$~million for the shell setup, $1.50$~million for the hexagonal
close-packed and WVT configurations).  The black line in the figure
indicates the analytic solution at this time.  All three simulations
used the same gamma-law equation of state, with $\gamma = 7/5$.

In the same way that initiating a Sedov blast wave calculation by
injecting energy into a single mesh cell could imprint the mesh
geometry onto the resulting shock, initiating a calculation by
injecting energy into a single SPH particle at the origin could
produce an aspherical explosion imprinted with artifacts of the
arrangement of neighboring particles.  In each of the three
simulations, energy $E = 1$ was injected into a small spherical volume
at the center of the simulation at $t = 0$.  The radius of that volume
varied among the simulations, according to the competing constraints
that it be as small as possible, to initiate a point-like explosion,
but large enough to extend out to several times the smoothing length
of the innermost particles, to eliminate relics of the specific
particle arrangement around the origin.

A uniform lattice is, by definition, poorly suited for problems with
spatially-varying resolution requirements.  The goal was to simulate a
Sedov blast wave in a sphere of radius $r_{max} = 1$, but the
hexagonal close-packed lattice compromised at both small and large
radii; with a uniform spacing of $0.01$ between closest neighbors, it
extended out only to $r_{max} = 0.63$.  At the same time, the uniform
lattice had limited ability to simulate a point-like explosion; energy
was smoothed over particles at radii $r < 0.024$, which included 81
particles.

Though both the concentric shell setup and the WVT setup covered a
larger range of radii, using the concentric shell setup for Sedov
simulations requires extra attention to the compromise between radial
and angular resolution.  Increasing the particle count for a given
shell tends to reduce variation in density around the sphere at that
radius, but spending the particle budget on angular resolution
requires reducing the overall number of shells, and having too few
shells per neighborhood can lead to radial fluctuations in density and
velocity.

Both the concentric shell setup and the WVT setup extended from
$r_{max} = 1$ (where the largest particles had smoothing lengths
$h_{max} = 2.62 \times 10^{-2}$ in the concentric shell setup and
$h_{max} = 3.29 \times 10^{-2}$ in the WVT setup) to much smaller
scales; near the origin, the smallest particles had smoothing lengths
$h_{min} = 7.81 \times 10^{-3}$ in the concentric shell setup and
$h_{min} = 2.69 \times 10^{-4}$ in the WVT setup.  In the concentric
shell setup, energy was smoothed over particles at radii $r < 3.9
\times 10^{-3}$, or the innermost 10 shells (containing a total of
$4,840$ particles); in the WVT setup, energy was smoothed over
particles at radii $r < 5.0 \times 10^{-3}$, which included $38,148$
particles.  Particles had an average of 54 neighbors in the hexagonal
close-packed lattice, 76 neighbors in the concentric shell setup, and
50 neighbors in the WVT setup.

As the shock expands outward, the variation of resolution with radius
among the three sets of initial conditions becomes apparent; at $t =
0.06317$, there are $481,562$ shocked particles in the simulation
using the shell setup, only $190,472$ in the hexagonal close-packed
simulation, and $977,512$ in the WVT simulation.

In the hexagonal closed-packed lattice, the shock propagates faster
along lattice planes than in other directions.  This leads to a radial
spread in the apparent shock location, averaged over the sphere, and
areas of lower density behind those advanced parts of the shock that
show up especially between radii $0.32 < r < 0.36$, at this time in
the simulation.  A WVT setup with the same uniform particle spacing
avoids this angle-dependent behavior, which eliminates the appearance
of faster-than-expected features ahead of the shock, and most of the
variation in density at a given radius.  When used to generate initial
conditions with spatially-varying resolution, WVT produces much less
scatter with the same total particle count, limiting the
numerically-seeded turbulence in this problem.

The nature of the concentric shell setup introduces a density
perturbation within each shell, visible as scatter in density at
discrete radii outside the shock in the right panel of
Figure~\ref{fig:sedov}.  This perturbation grows when the shock passes
through it, driving strong density perturbations and convection.  In
simulations of supernova explosions~\citep[e.g.,][]{FryerMixing} and
other more complex environments, ensuring that these perturbations are
small compared to perturbations expected in the natural system can
require several rounds of setup, simulation, and adjustment.  Even
with the same innermost radius and particle size, WVT produces a
smoother representation of the initial conditions in both radius and
angle; in the simulations presented here, WVT was used to produce a
configuration with a smooth representation at even smaller radii,
supporting a smaller energy injection region containing more
particles.

The WVT setup underestimates the density both behind and ahead of the
shock---by $6.6$\% for particles between $r = 0.2$ and $0.25$, $9.3$\%
between $r = 0.25$ and $0.3$, $15.4$\% between $r = 0.3$ and the
shock, and $2.9$\% in the unshocked region between $r = 0.37$ and
$0.41$.  Both the hexagonal closed-packed and shell configurations
provide better estimates of the density ahead of the
shock---overestimating by $0.95$\% and $1.9$\%, respectively.  Behind
the shock, the average density for the shell configuration is
consistently lower than the WVT result---$15.5$\% below the analytic
value for particles between $r= 0.2$ and $0.25$, $14.4$\% low between
$r = 0.25$ and $0.3$, and $13.6$\% low between $r = 0.3$ and the
shock.  The average density in the hexagonal close-packed simulation
is just slightly above the analytic line---by $0.6$\%---for particles
between $r = 0.2$ and $0.25$, and below the target value by $6.7$\%
between $r = 0.25$ and $0.3$, and by $14.7$\% between $r = 0.3$ and
the shock, but averaging over all particles at a given radius in this
calculation hides significant variation between different angles
around the sphere.  For both the shell setup and WVT, the iterative
process of assigning masses to particles given their initial position
and spacing could be improved to better match the desired initial
density profile.


\section{Comparison Tests}
\label{s.comparison}

\subsection{Interpolation Accuracy: Uniform Density}

An important performance test for any SPH setup method is to find out
how well it reproduces a given density field. This test will reveal
the level of perturbations that are introduced by the setup, which
could seed convection, excite sound waves or trigger
instabilities. The simplest such test is to see how well each method
can mimic a uniform density field with a uniform particle
distribution. Thus, each particle should have the same mass, which we
will assume to be $1$, and the same smoothing length/resolution. At
the same time, this test will then provide a means to test the
accuracy of the particle density distribution itself.

\begin{figure*}
  \includegraphics[width=0.195\textwidth]{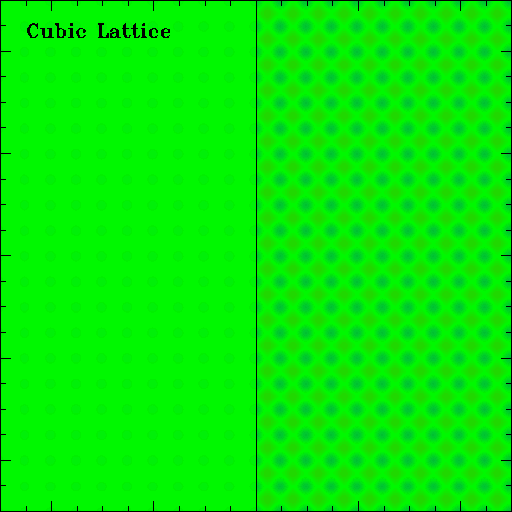}
  \includegraphics[width=0.195\textwidth]{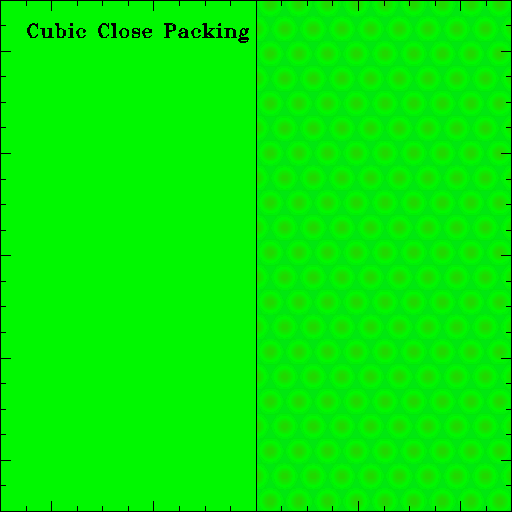}
  \includegraphics[width=0.195\textwidth]{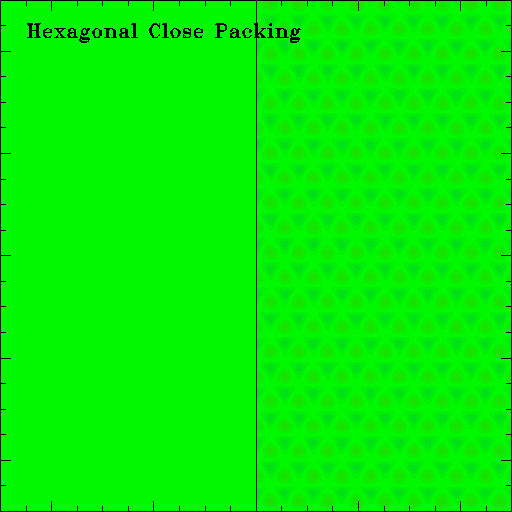}
  \includegraphics[width=0.195\textwidth]{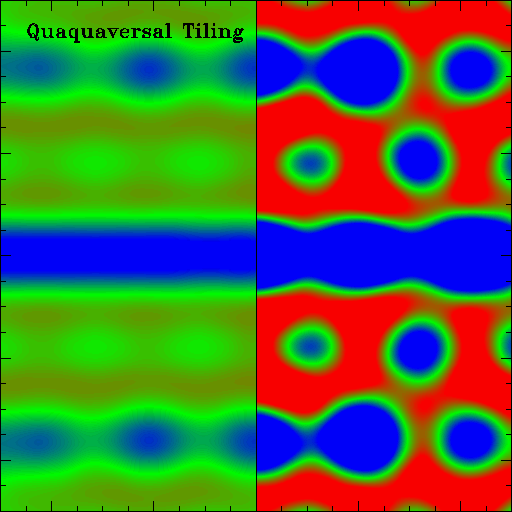}
  \includegraphics[width=0.195\textwidth]{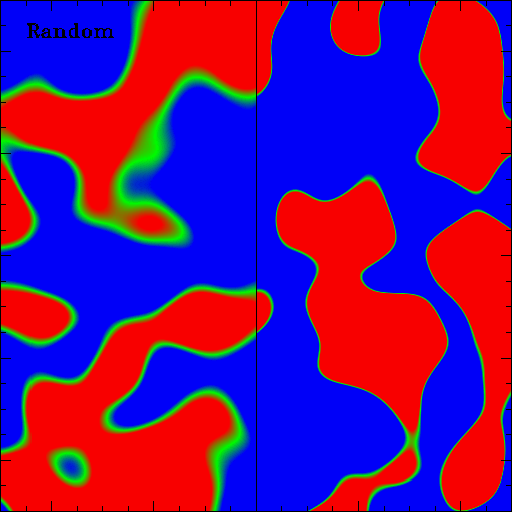}
  \null\hspace{0.195\textwidth}
  \includegraphics[width=0.195\textwidth]{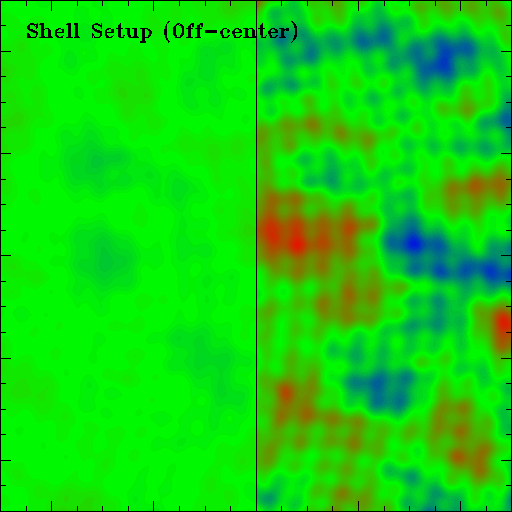}
  \includegraphics[width=0.195\textwidth]{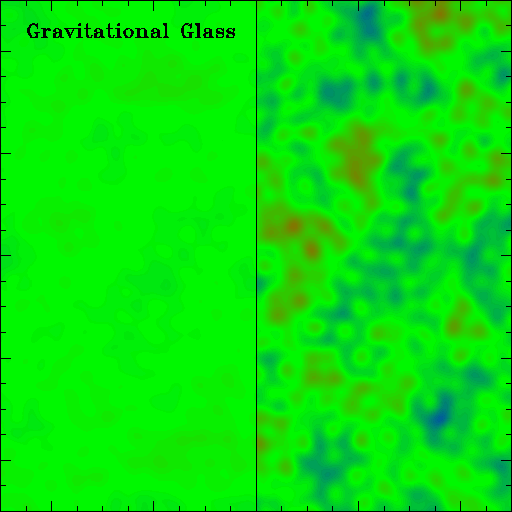}
  \includegraphics[width=0.195\textwidth]{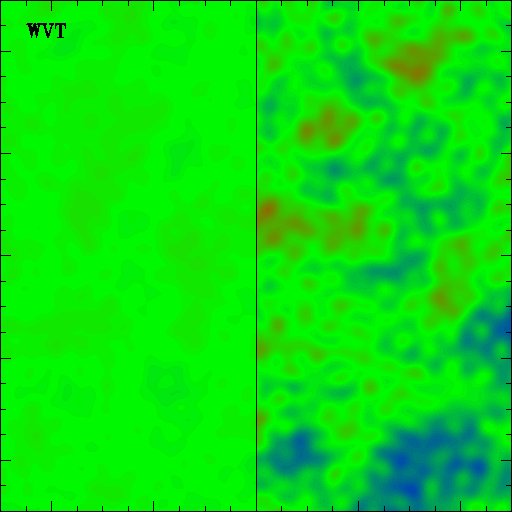}
  \hspace{.72cm}
  \includegraphics[angle=90, width=0.048\textwidth]{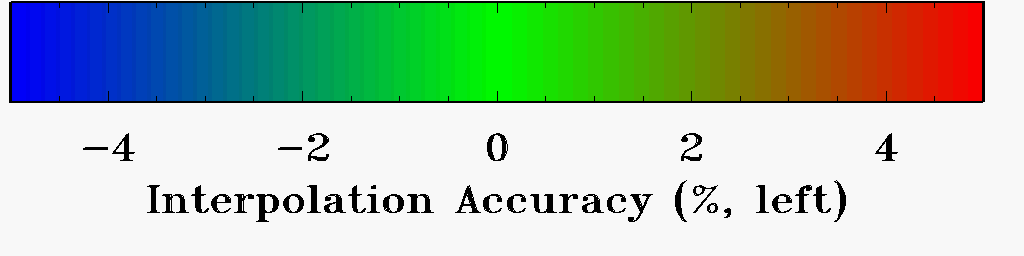}
  \hspace{.6cm}
  \includegraphics[angle=90, width=0.048\textwidth]{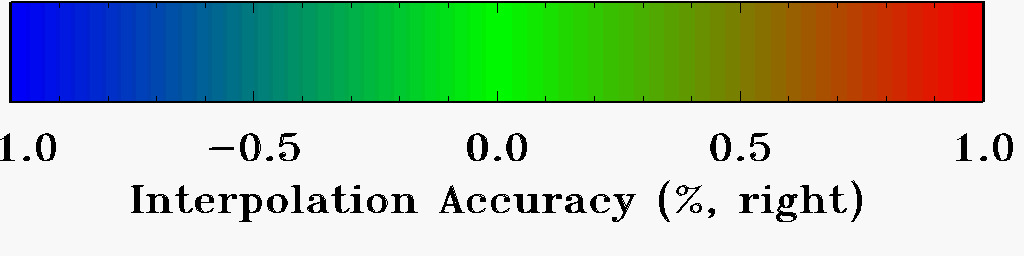}
  \caption{Comparison of the interpolation accuracy for 128 neighbors
    in the cubic lattice, cubic close packing, hexagonal close
    packing, quaquaversal tiling, random configuration, shell setup,
    gravitational glass, and the new WVT approach (top left to bottom
    right). Colors indicate deviations from the target density, with
    blue colors showing negative and red colors denoting positive
    deviations. Each panel is divided into two halves with different
    dynamic ranges: $\pm5\%$ on the left and $\pm1\%$ on the right
    side. Note that the shell setup is shown at an off-center
    location, to avoid the discussed special treatment of the center
    in this comparison. Quaquaversal tiling and the random setup
    perform noticeably worse than any other method, while the uniform
    grid setups perform best as expected. The non-gridded setup
    methods perform equally well, with very high interpolation
    accuracies that never exceed $1\%$.}
  \label{f.uniformrho}
\end{figure*}

Figure~\ref{f.uniformrho} shows the accuracy of all uniform density
methods described in \S\ref{s.methods}, along with the new WVT
method. Each panel shows a projection of a unit cube containing 8,000
particles onto the x--y plane according to the standard spline SPH
kernel targeted at containing approximately 128 neighbors. We divided
each figure into two parts, with the colors in the left half showing
up to $5\%$ deviations (negative: blue, positive: red, accurate:
green), whereas the right half reveals lower level (up to $1\%$)
deviations.

The first three panels show the lattice configurations (cubic lattice,
cubic close packing, and hexagonal close packing) which obviously have
excellent interpolation properties. This is not surprising, as they
are designed to be as uniform as possible, and each particle has an
identical number of neighbors. Only in the right half of each panel
does low-level noise becomes visible, revealing the underlying lattice
structures. Note that the cubic close packing panel shows hexagonal
structures, as the x--y plane cuts through a hexagonal layer. Other
orientations of the plane would reveal different patterns, but give
the same qualitative impression.

The fourth panel in the top row shows the quaquaversal tiling
configuration, which demonstrates very strong clustering of particles
and corresponding deviations from the ideal values. Even with 128
neighbors, density fluctuations on the order of $5\%$ are found
throughout the simulation volume. In addition, these deviations are
strongly spatially correlated, which makes numerical artifacts
likely. This reason alone is grounds enough not to use quaquaversal
tiling for SPH setups. This effect has also recently been pointed out
by \citet{WhiteNbodyIC}, who found that using quaquaversal tiling to
initialize cosmological situations leads to an excessive amount of
small halos and clumping. In fact, the only setup method that produces
stronger density fluctuations is placement of particles randomly
throughout the simulation domain (top right panel).

The first panel in the lower row shows the interpolation accuracy of
the shell setup, at an off-center location.  This setup has most often
been used for simulations with an inner boundary inside the innermost
shell of particles, which excludes that central volume from the
simulation domain.  (In a simulation that extends all the way to
$r=0$, interpolation accuracy inside the innermost shell will be poor
without special treatment.  Options include placing a small lattice
configuration, a gravitational glass configuration, or a single
particle inside the innermost shell; the former two can still produce
artifacts near the innermost shell, while the latter works best if the
innermost shell radius is about the same as the typical inter-particle
separation within the shell.)  Similar to the lattice configurations,
individual shells are visible as low-level noise in the right half of
the panel. The level of noise in the density interpolation for 128
neighbors is on the order of $1\%$, consistent with findings by
\citet{FryerMixing}. However, note that the setup is only optimized
within one shell, which leads to the noise deviations having a
preferred radial direction perpendicular to the shell structures.

Within one shell, the setup has similar properties to a uniform
gravitational glass, shown in the second panel of the bottom row. The
level of noise for this method is very low, generally on the order of
$0.5\%$ at most. Also note that the noise is isotropic with no
preferred direction, as is true for the underlying particle
distribution. These desirable properties make the gravitational glass
setup a suitable choice for uniform density distributions.

The last panel in the bottom row shows the interpolation properties of
our new WVT setup method. Note the similarity to the gravitational
glass, with an equally low amount of noise and an isotropic
distribution of the noise without preferred directions. In our WVT
implementation, each repulsive force is roughly proportional to
$r^{-2}$, and the ratio of scale lengths is $1$ for a uniform
distribution, which makes the WVT setup locally very similar to a
gravitational glass.

In summary, the three uniform lattices (cubic lattice, cubic close
packing, hexagonal close packing) have excellent interpolation
characteristics for a uniform density and should be used in situations
where lattice effects are expected to be unimportant, but a very low
level of initial numerical noise is needed. The quaquaversal and
random initial condition are unacceptable for any application due to
their low interpolation accuracy. In spherical symmetry, the shell
setup provides an adequate configuration, but without a special
treatment of the center is unable to reproduce a solid, uniform center
of the sphere. The gravitational glass and the new WVT setup method
both perform best in the interpolation test among non-lattice
configurations. Both have a high level of interpolation accuracy with
maximum deviations generally on the order of $0.5\%$ for 128
neighbors.

\subsection{Interpolation Accuracy: Non-Uniform Density}

\begin{figure*}
  \includegraphics[width=0.195\textwidth]{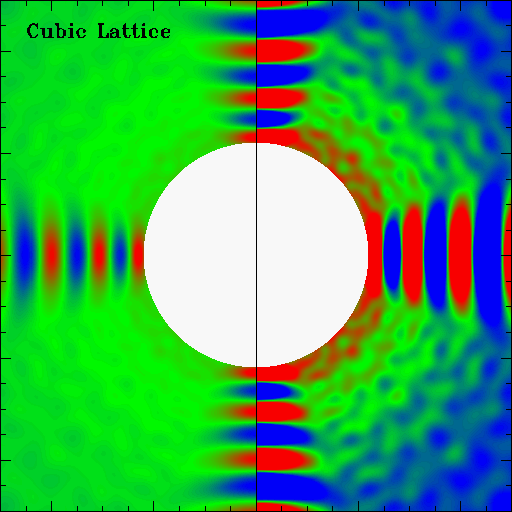}
  \includegraphics[width=0.195\textwidth]{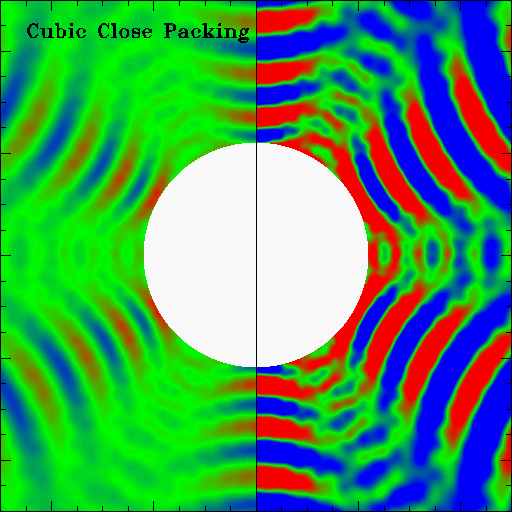}
  \includegraphics[width=0.195\textwidth]{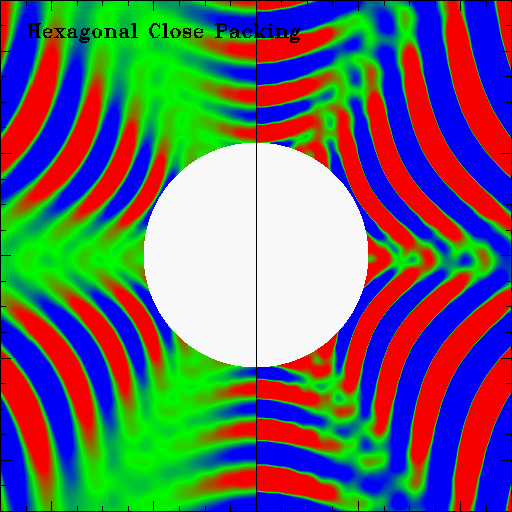}
  \includegraphics[width=0.195\textwidth]{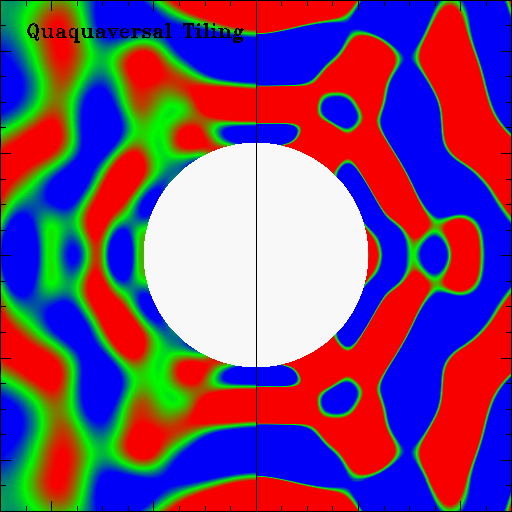}
  \includegraphics[width=0.195\textwidth]{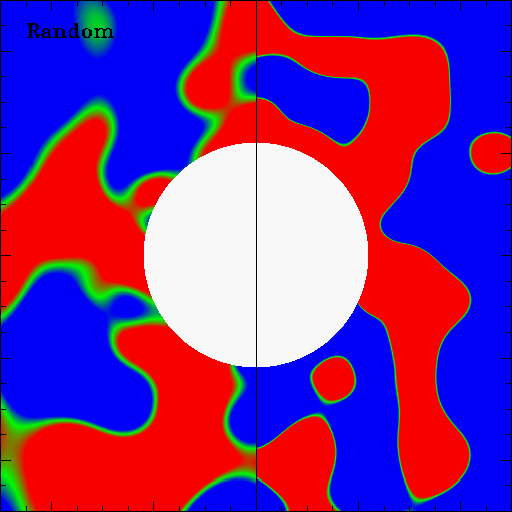}
  \null\hspace{0.195\textwidth}
  \includegraphics[width=0.195\textwidth]{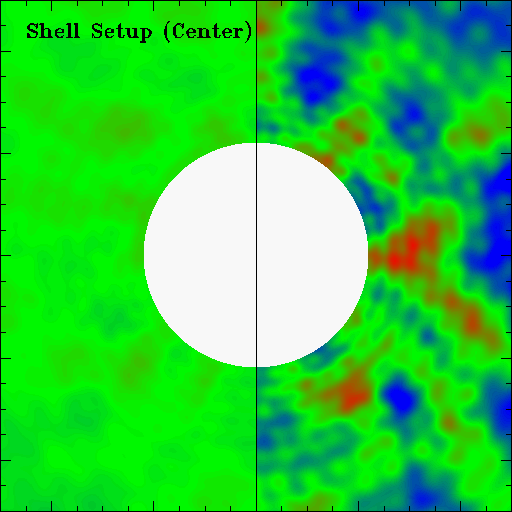}
  \includegraphics[width=0.195\textwidth]{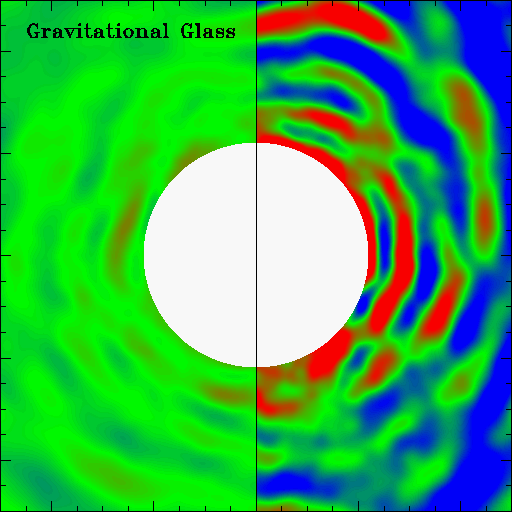}
  \includegraphics[width=0.195\textwidth]{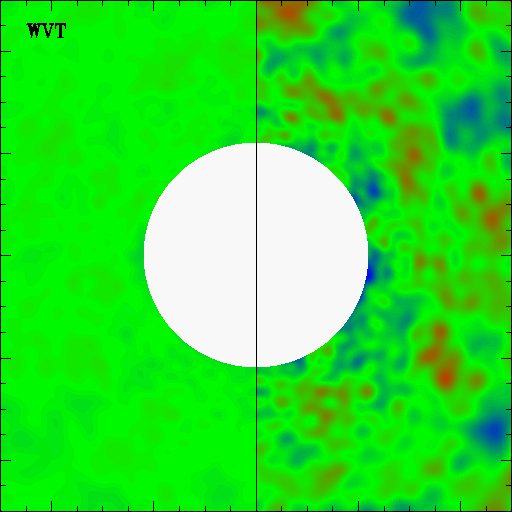}
  \hspace{.72cm}
  \includegraphics[angle=90, width=0.048\textwidth]{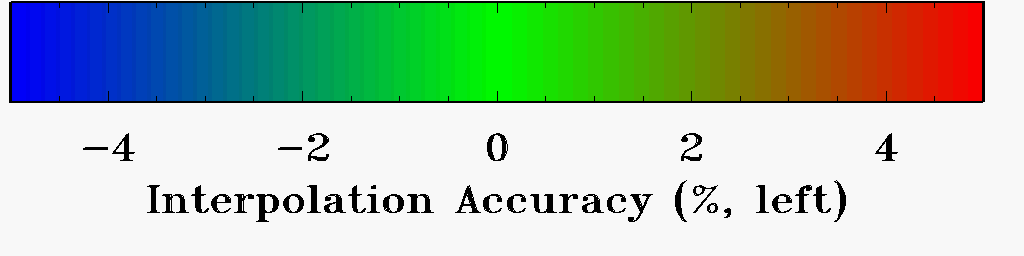}
  \hspace{.6cm}
  \includegraphics[angle=90, width=0.048\textwidth]{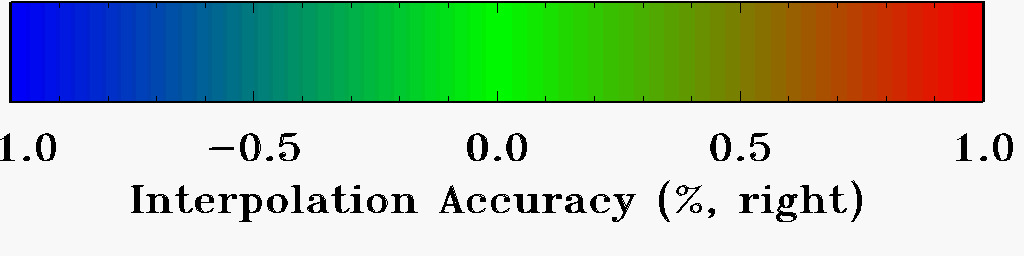}
  \caption{The equivalent of Figure \ref{f.uniformrho}, but for
    spatially adaptive setups: stretched cubic lattice, stretched
    cubic close packing, stretched hexagonal close packing, stretched
    quaquaversal tiling, random configuration, shell setup, stretched
    gravitational glass, and the new WVT approach (top left to bottom
    right). All uniform-grid-based setups, quaquaversal tiling and the
    random configuration perform very badly. Of the non-gridded setup
    methods, the WVT setup performs best with density inaccuracies
    below $1\%$. The stretched gravitational glass introduces
    artifacts that are located inside shells, whereas the shell setup
    demonstrates deviations in the radial direction.}
  \label{f.stretchedrho}
\end{figure*}

We now consider a very similar test, but for a non-uniform particle
distribution. We test all adaptive setup methods listed in
\S\ref{s.methods} along with our new WVT setup. As a comparison test,
we choose a spherical setup with the resolution intended to scale as
$r^{2/3}$. Figure~\ref{f.stretchedrho} shows the comparison data, in a
similar fashion to Figure~\ref{f.uniformrho}.

Even with this moderate amount of stretching, all the lattice
configurations (stretched cubic lattice, stretched cubic close
packing, stretched hexagonal close packing) perform very poorly in
this test. This is not surprising when one considers the 3D structure
of the stretched lattices in Figure \ref{f.popularconf_adaptive}. The
problem for these structures is that the stretching factor is a
function of radius, and thus not parallel to one of the lattice
axes. Thus, originally parallel planes are warped significantly during
the stretch process, and the particle spacings within one such plane
are multiplied by different stretching factors. This results in a very
uneven distribution of particle density, and the lattice structure can
now be easily picked out in the first three panels of
Figure~\ref{f.stretchedrho}.

The quaquaversal tiling and random configuration perform even more
poorly, and the problems seen in the uniform density test are even
more apparent.

The intrinsically adaptive shell setup (bottom row, left) performs
very well in this test, with density perturbations equivalent to the
uniform density test, usually not exceeding $1\%$ for 128
neighbors. The only disadvantage of the shell setup is that the
density deviations are systematically in the radial direction, as it
is only optimized within a shell.

\begin{figure*}
  \begin{center}
    \includegraphics[width=\textwidth]{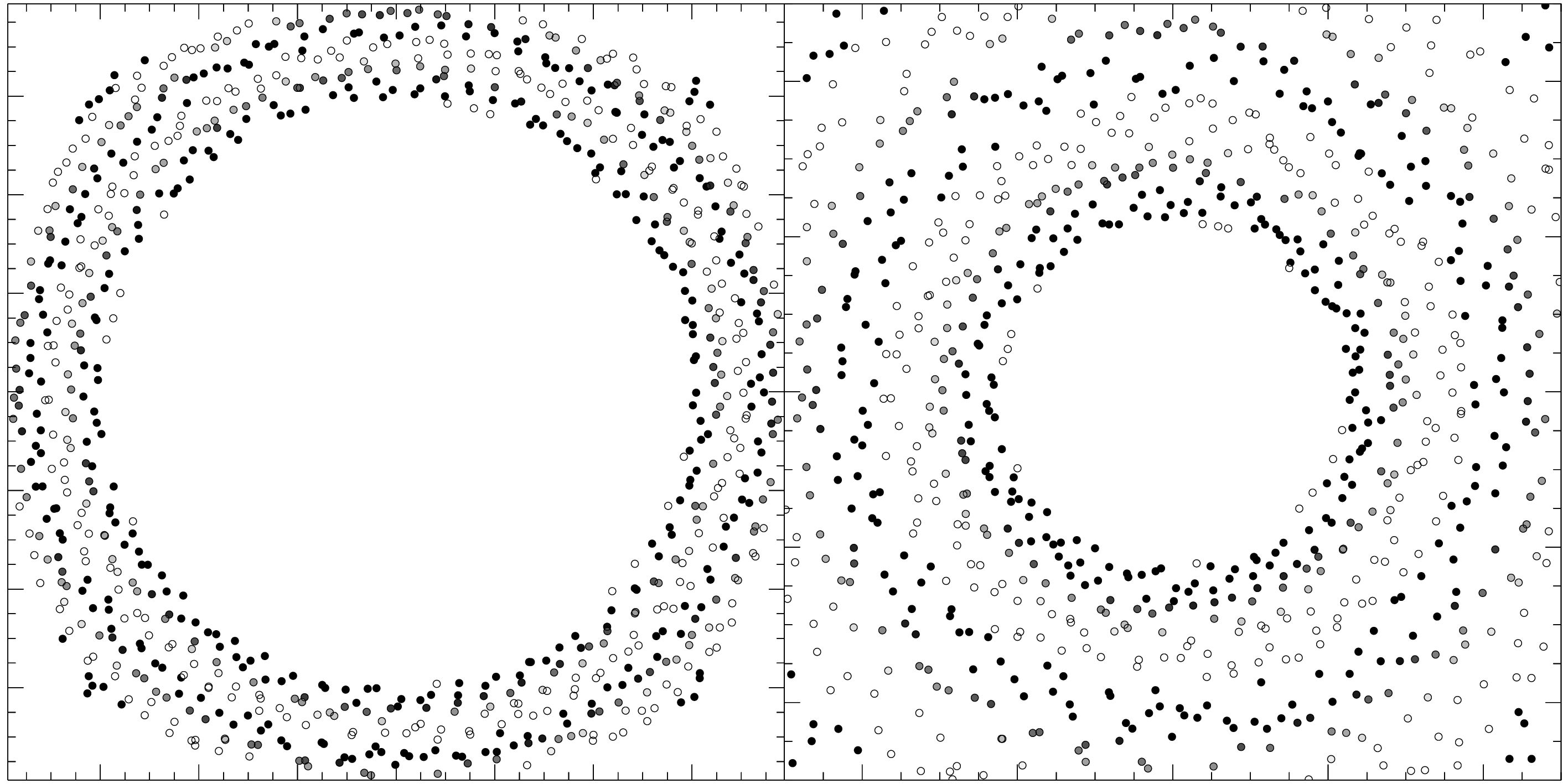}
  \end{center}
  \caption{Unstretched (left) and stretched glass (right). Note the
    wavy shell-like structure in the stretched glass. To bring out
    this structure, particles are periodically colored by their radius
    in the unstretched glass, which is then also applied to the
    stretched glass.}
  \label{f.glass}
\end{figure*}

Interestingly and maybe surprisingly, the stretched glass performs
relatively poorly, as shown in the middle panel of the bottom
row. Even though the gravitational glass has excellent interpolation
properties for uniform densities, this is not the case when stretched
in a radial (or any other) direction. As was true for the lattice
configurations, the stretching procedure tends to pronounce voids
between planes that are perpendicular to the direction in which the
stretching is applied. The glass does not have a uniform clear lattice
structure but still tends to order particles along randomly oriented
strings on a local level, as can be seen in the left panel of
Figure~\ref{f.glass}. Thus the stretching will preferably pick out
those features that are perpendicular to our stretching direction,
i.e. those inside shells. This leads to the wavy shell-like features
in the right panel of Figure~\ref{f.glass} which shows the stretched
glass. In this particular example, the density deviations are on the
order of $3\%$, which is significantly less than the lattice
structures. However, with stronger stretching, these features will
only become even more pronounced.

The WVT setup (bottom row, far right panel) does not exhibit these
features. Note how the level of noise for the adaptive setup is as low
as for the uniform distribution. This is due to the fact that the WVT
setup knows beforehand the desired resolution at each point in space,
and is not a stretched version of a uniform distribution, allowing it
to converge to the optimal solution in either case. We also note that
this high level of interpolation accuracy does not depend on spherical
symmetry as in the shell setup, which makes WVT much more versatile.

In summary, we find that stretched lattice configurations are not
suitable for producing non-uniform particle distributions.
Quaquaversal tiling and random configuration have even worse
interpolation properties. The shell setup has acceptable levels of
noise, but is restricted to spherical symmetry and the noise
distribution is preferably along the radial direction. The stretched
glass setup also introduces artifacts in shells which could lead to
radial pulsations in an SPH simulation. The WVT setup has
interpolation properties that are equivalent to the uniform
distribution, which makes it the method of choice for adaptive
resolution requirements. All other distributions should be relaxed
into their equilibrium configuration prior to being used.

\subsection{Particle Noise}

Another way of judging the characteristics of an SPH setup is to
measure the particle noise inside a uniform distribution of particles
within a uniform density distribution. In an ideal situation, all the
pressure forces of individual particles cancel out, and the net force
is 0 or very small compared to an individual force. Then, the
contribution to the pressure force of an individual SPH particle $i$
on another particle $j$ is \citep{MonaghanSPH}
\begin{equation}
  {dv_{ij} \over dt}\propto -m_i {P_i \over \rho_i^2} \nabla_i W_{ij}.
\end{equation}

For our test, we set up conditions such that the particle masses
$m_i=1$, the average pressure $\bar P_i=1$ and the average density
$\bar\rho_i=1$. Thus, we expect a single particle to contribute
$|{\nabla W_{ij}}|\approx 0.08$ on average to the acceleration
term. Assuming Poisson noise, we thus expect the noise for a random
configuration to be on the order of $(N_{\rm neigh})^{1/2}\, |{\nabla
  W_{ij}}|=0.64$ for 64 neighbors.\footnote{Although this relationship
  suggests that increasing $N_{\rm neigh}$---by, implicitly,
  increasing $h$---would conveniently reduce noise, \citet{price12}
  notes that such stretching amounts to an arbitrary change in the
  weighting of a neighboring particle at a given distance and does not
  lead to formal convergence of the density estimate, while using
  higher-order spline kernels improves the smoothness of the density
  estimate without changing the definition of $h$ and should be the
  standard approach.}  Our random configuration yields a value of
$0.60\pm0.25$ in our test setup, consistent with expectations. We will
take this measured value of 0.6 as a reference point to measure the
performance of the other setup methods in terms of ``fractional
Poisson noise''.

As expected, the uniform lattice configurations yield a perfect
equilibrium down to machine precision. The quaquaversal tiling on the
other hand shows very poor performance again, with a particle noise of
about $30\%$ the Poisson level, consistent with our findings in the
density interpolation accuracy. Both the shell setup ($3.9\pm1.7\%$),
the gravitational glass ($3.1\pm1.3\%$) and WVT ($3.9\pm1.7\%$) reduce
the noise by an order of magnitude.

\subsection{Summary}

In Table~\ref{t.comparison}, we summarize our findings about positive
and negative characteristics of all considered setup methods for
uniform particle distributions. We also provide recommendations for
which method should be preferred in which situation.

\begin{table*}
  \caption{Comparison of Particle Setup Methods.}
  \label{t.comparison}
  \vbox to220mm{\vfil Landscape Table from external file table1.tex to
    go here.  \vfil}
\end{table*}


\section{Conclusions}
\label{s.summary}

We have presented an extensive comparison of all particle setup
methods currently employed in astrophysics that we are aware of. In
particular, we review spatially uniform configurations such as a cubic
lattice, cubic close packing, hexagonal close packing, quaquaversal
tiling, and gravitational glasses. For spatially adaptive methods, we
also include the random probability distribution, stretched lattice,
stretched glass, and a concentrical shell setup. To the best of our
knowledge, the stretched glass and concentrical shell setup have not
been described in the literature before.

The main focus of our paper, however, is a new setup method based on
weighted Voronoi tesselations. This new method allows for arbitrary
spatial configurations of particles, which has not been possible
before. We show that this new method is easy to implement on existing
SPH codes and demonstrate its superior characteristics in several
examples.

This method has now been used in a variety of astrophysics problems
from core-collapse supernovae~\citep{Ellinger11} to modeling binary
interactions~\citep{RaskinWD1, RaskinWD2, FryerWDmerger}.  Especially
in doing binary interaction calculations comparing multiple
techniques, using identical initial conditions is critical and these
initial conditions tend to have complicated structures caused by tidal
effects~\citep[e.g.,][]{Passy11}.


\section*{Acknowledgments}

This work was carried out under the auspices of the National Nuclear
Security Administration of the U.S. Department of Energy at Los Alamos
National Laboratory under Contract No. DE-AC52-06NA25396.
Three-dimensional images were created using the \texttt{VisIt} package
developed at LLNL.


\end{document}


\begin{landscape}
\setcounter{table}{0}\begin{table*}\begin{minipage}{\textheight}
\caption{Comparison of SPH Particle Setup Methods.}
\label{t.comparison}
\begin{tabular}{@{} p{0.1\textwidth} c p{0.05\textwidth} p{0.355\textwidth} p{0.355\textwidth} p{0.355\textwidth}}
\hline
{\bf Method} & {\bf Adaptive?} & {\bf Noise} & {\bf Advantages} & {\bf Disadvantages} & {\bf Recommendation}\\
\hline
\hline
{\bf Cubic Lattice} & no 
& 
very low 
& 
- very good interpolation properties \newline
- trivial to implement \newline
& 
- prone to artifacts due to very strong particle alignment along grid axes \newline 
- uniform distributions only \newline
- unstable equilibrium configuration \newline
- not an optimal packing scheme 
&
- Should not be used, as it is affected by strong lattice effects and does not present a stable equilibrium configuration\\
\hline
{\bf Cubic Close Packing} 
&
no 
& 
very low 
&
- optimal packing scheme \newline
- stable equilibrium configuration
&
- uniform distributions only\newline
- prone to artifacts due to strong particle alignment along lattice symmetry axes 
 &
- Useful for mainly static problems that do not collapse or exhibit shocks, and if interpolation accuracy is important\\
\hline
{\bf Hexagonal Close Packing} 
& 
no 
& 
very low 
&
- optimal packing scheme \newline
- stable equilibrium configuration
&
- uniform distributions only\newline
- prone to artifacts due to strong particle alignment along lattice symmetry axes 
&
- Useful for mainly static problems that do not collapse or exhibit shocks, and if interpolation accuracy is important\\
\hline
{\bf Quaquaversal Tiling} 
& 
no 
& 
high 
& 

&
- non-intuitive lattice effects \newline
- uniform distributions only
& 
- Not recommended for use with SPH\\
\hline
{\bf Random Configuration} & yes & very high 
&
- applicable to arbitrary geometries 
& 
- very bad interpolation properties\newline
- very clumpy\newline
- dominated by Poisson noise
&
- Not recommended for use with SPH\\
\hline
{\bf Gravitational Glass} & no 
& low 
&
- very good interpolation properties\newline
- no lattice structures\newline
- with periodic boundaries, can be tiled to generate arbitrarily large setups \newline
- noise is isotropic
&
- uniform distributions only
& 
- A very good choice for uniform particle distributions if interpolation accuracy is important and lattice effects need to be avoided\\
\hline
{\bf Stretched Lattice} & yes
& 
very high
&
- fast and easy to implement
& 
- strongly anisotropic particle distribution due to lattice effects\newline
- not a stable configuration, has to be relaxed into equilibrium configuration
&
- Should only be used in conjunction with a relaxation technique\\
\hline
{\bf Stretched Glass} & yes
& 
high
&
- better starting point for initial conditions than a stretched lattice
&
- noise non-isotropic, tends to be spatially correlated in shells.\newline
- not a stable configuration, has to be relaxed into equilibrium configuration
&
- Should only be used in conjunction with a relaxation technique; faster to relax than a stretched lattice\\
\hline
{\bf Shell Setup} & yes
& 
low 
&
- very well suited to produce perfect spherical symmetry \newline
- very good interpolation properties
&
- computationally expensive to generate shell templates\newline
- requires a lot of shell templates \newline
- only optimized within shells \newline
- noise non-isotropic
&
- Good choice for spherically-symmetric initial conditions if interpolations accuracy is important. Center needs special treatment. \\
\hline
{\bf WVT Setup} & yes 
& 
low  
& 
- applicable to arbitrary geometries \newline
- easy to implement on top of an existing SPH code \newline
- very good interpolation properties
& 

&
- Good choice for any uniform or adaptive setup. Only choice for adaptive setups without spherical symmetry. \\
\hline
\end{tabular}
\end{minipage}
\end{table*}
\end{landscape}

